\documentclass[draftcls, 12pt,onecolumn,oneside]{IEEEtran}
\usepackage{amsmath,amsfonts}
\usepackage{array}
\usepackage[caption=false,font=scriptsize,labelfont=rm,textfont=rm]{subfig}
\usepackage{textcomp}
\usepackage{stfloats}
\usepackage{url}
\usepackage{verbatim}
\usepackage{graphicx}
\usepackage{cite}
\makeatletter
\newcommand{\rmnum}[1]{\romannumeral #1}
\newcommand{\Rmnum}[1]{\expandafter\@slowromancap\romannumeral #1@}
\makeatother

\usepackage{CJK}
\usepackage{indentfirst}
\usepackage{cases}

\usepackage{amssymb}
\usepackage{enumerate}
\usepackage{graphicx}
\usepackage{amsmath}
\usepackage{bm}
\usepackage{amsfonts}
\usepackage{amsmath}
\usepackage{algorithmic}
\usepackage[ruled]{algorithm2e}
\usepackage{textcomp}
\usepackage{color,xcolor}
\usepackage{indentfirst}
\usepackage{amsthm}
\usepackage{epstopdf}
\usepackage{boxedminipage}

\allowdisplaybreaks[3]

\newtheorem{Lemma}{Lemma}
\newtheorem{mypro}{Proposition}
\newtheorem{remark}{Remark}
\begin{document}
	%
	\title{Communication-Sensing Region for Cell-Free  Massive MIMO ISAC Systems}
	
	\author{Weihao Mao, \IEEEmembership{Student Member},~Yang Lu,~\IEEEmembership{Member,~IEEE},\\
		 Chong-Yung Chi,~\IEEEmembership{Life Fellow,~IEEE},  Bo Ai, \IEEEmembership{Fellow, IEEE}, \\
		 Zhangdui Zhong, \IEEEmembership{Fellow, IEEE},~{\rm and}~Zhiguo Ding,~\IEEEmembership{Fellow,~IEEE}
	\thanks{Weihao Mao and Yang Lu are with the School of Computer and Information Technology, Beijing Jiaotong University, Beijing 100044, China (e-mail: 21120382,yanglu@bjtu.edu.cn).}
	\thanks{Chong-Yung Chi is with the Institute of Communications Engineering, Department of Electrical Engineering, National Tsing Hua University, Hsinchu 30013, Taiwan (e-mail:cychi@ee.nthu.edu.tw).}
	\thanks{Bo Ai and Zhangdui Zhong are with the State Key Laboratory of Rail Traffic Control and Safety, Beijing Jiaotong University, Beijing 100044, China, and Bo Ai is also with the School of Electronics and Information Engineering, Beijing Jiaotong University, Beijing 100044, China (e-mail: boai,zhdzhong@bjtu.edu.cn).}
	\thanks{Zhiguo Ding is with Department of Electrical Engineering and Computer Science, Khalifa University, Abu Dhabi, UAE, and Department of Electrical and Electronic Engineering, University of Manchester, Manchester, UK (e-mail: zhiguo.ding@manchester.ac.uk).}
}

	\maketitle
		
	\vspace{-3em}
	
	\begin{abstract}
		This paper investigates the system model and the transmit beamforming design for the Cell-Free massive multi-input multi-output (MIMO) integrated sensing and communication (ISAC) system. The impact of the uncertainty of the target locations on the propagation of wireless signals is considered during both uplink and downlink phases, and especially, the main statistics of the MIMO channel estimation error are theoretically derived in the closed-form fashion. A fundamental performance metric, termed communication-sensing (C-S) region, is defined for the considered system via three cases, i.e., the sensing-only case, the communication-only case and the ISAC case. The transmit beamforming design problems for the three cases are respectively carried out through different reformulations, e.g., the Lagrangian dual transform and the quadratic fractional transform, and some combinations of the block coordinate descent method and the successive convex approximation method. Numerical results present a 3-dimensional C-S region with a dynamic number of access points to illustrate the trade-off between communication and radar sensing. The advantage for radar sensing of the Cell-Free massive MIMO system is also studied via a comparison with the traditional cellular system. Finally, the efficacy of the proposed beamforming scheme is validated in comparison with zero-forcing and maximum ratio transmission schemes.
	\end{abstract}

	%
	\IEEEpeerreviewmaketitle
	\setlength{\parindent}{1em}
	\section{Introduction}
	 \vspace{-1em}
	\subsection{Background}
	With the development of mobile intelligent applications, e.g., auto-driving and indoor location, sensing has played a more important role than ever before, which motivates the joint design of the communication and sensing \cite{bc_isac1}.
	Integrated sensing and communication (ISAC) is an emerging technique which utilizes the same spectrum resource to realize both communication and radar sensing functions \cite{bc_isac2}. In the literature, there are two ISAC architectures. One is the coexisting radar and communication (CRC) architecture, where the radar and the communication base station (BS) are located separately, and the sensing beam and the communication beam are also generated individually \cite{bc_isac3}. The other is the dual functional radar-communication (DFRC) architecture, where the radar and the communication BS are co-located, i.e., they are integrated as an ISAC BS with shared hardwares \cite{bc_isac4}. Therefore, the sensing beam and the communication beam can be jointly generated. Compared with the CRC architecture, the DFRC architecture can be more cost-efficient and has been regarded as one of the most promising technical directions in B5G era \cite{bc_isac5}.
	
	Meanwhile, the massive multi-input multi-output (MIMO) technology has been proposed as a spectral efficiency (SE) enabler, which facilitates that multiple users share the same time-frequency resource with limited inter-user interference \cite{bc_massive_mimo2}.  An emerging massive MIMO architecture, named Cell-Free massive MIMO \cite{bc_cell-free1}, draws increasing attention, where all antennas which would be collocated on a single traditional BS are geographically distributed over multiple access points (APs) which are connected to a central processing unit (CPU). The Cell-Free massive MIMO system has been shown effective to significantly improve the median wireless coverage performance including both SE and energy efficiency (EE) \cite{bc_cell-free2}. Besides, the massive MIMO technology is capable of providing high resolution radar sensing \cite{bc_massive_mimo4}. Therefore, it is attractive to integrate the Cell-Free massive MIMO and the ISAC into a single system, termed the Cell-Free massive MIMO ISAC system.
	
	On the other hand, a trade-off exists in the ISAC system as the two functions for communication and radar sensing share the available resources, including spectrum bandwidth, power budget and transmit antennas (referred to as spatial degree of freedom) et al. Most existing works concentrated on the communication/radar sensing performance under respective design requirements. The former is termed as the communication-centric design and the latter as the sensing-centric design. Nevertheless, few existing works focused on the study of the entire operation region of the dual functionalities. Besides, in \cite{bc_trade}, it was verified that by utilizing some wireless coverage enablers, the trade-off can be improved. In view of the advantages of the Cell-Free massive MIMO, it is worthwhile to further explore the trade-off in Cell-Free massive MIMO ISAC systems.
	
	\vspace{-1em}
	\subsection{Related Works}  
	 There have been extensive existing works on ISAC systems, see e.g., \cite{crc1, crc2, dfrc_r1, dfrc_r2, dfrc_c1, dfrc_c2, sc1, sc2}. In \cite{crc1}, the radar signal-to-interference-plus-noise ratio (SINR) was defined and maximized under the constraints of the power budget and the quality of service (QoS) requirements of users for a CRC system. In \cite{crc2}, a transmit design was investigated with the aim to maximize the communication SINR for a double reconfigurable intelligent surfaces (RISs)-assisted CRC system, and it was illustrated that with the help of RISs, the communication performance can be improved while satisfying the radar sensing requirement. To reduce the hardware costs in ISAC systems, some works adopted the DFRC architecture. In \cite{dfrc_r1} and \cite{dfrc_r2}, the sensing-centric hybrid beamforming design was studied for DFRC systems to maximize the sensing performance under the QoS requirements, where the beampattern matching mean square error (MSE) and the Cramér-Rao bound (CRB) were used as the sensing criteria, respectively. Apart from sensing-centric designs, some works considered the communication-centric designs in DFRC systems. In \cite{dfrc_c1}, a RIS-assisted single-user DFRC system was investigated, where the user's signal-to-noise ratio (SNR) was maximized under the detection probability constraint of radar sensing. In \cite{dfrc_c2}, the sum secrecy rate of the network was maximized under the constraint of the radar SNR for a non-orthogonal multiple access (NOMA)-assisted DFRC system. However, the trade-off between communication and radar sensing in ISAC has not yet been reported comprehensively in above mentioned designs. In \cite{sc1}, the communication-sensing (C-S) region was defined and derived for a single-target DFRC system, where the sum rate and the signal-clutter-noise ratio (SCNR) were respectively utilized as the performance metrics for communication and sensing. In \cite{sc2}, the C-S region was obtained for a rate-splitting multiple access-assisted DFRC system  where the sum rate and the squared position error bound (SPEB) were respectively employed as the performance metrics for communication and sensing.
	 
	 The system models in \cite{crc1, crc2, dfrc_r1, dfrc_r2, dfrc_c1, dfrc_c2, sc1, sc2} were based on the traditional cellular architecture. Recently, Cell-Free massive MIMO has been regarded  as a promising candidate for the future mobile communication thanks to its superior coverage and high SE. The channel hardening effect in massive MIMO systems relies on the available statistics of the estimated channel \cite{cf_u1}, thereby making the pilot training of foremost importance. Due to the massive separately distributed APs, the pilot training is challenging in the Cell-Free massive MIMO system \cite{cf_u2}. The joint pilot training and information transmission designs have been investigated in many existing works on Cell-Free massive MIMO, see, e.g. \cite{cf_p1, cf_mrt1, cf_mrt2, cf_zf}. In \cite{cf_p1}, a channel estimation approach was proposed for a RIS-assisted Cell-Free massive MIMO system with spatially-correlated channels, together with its achievable communication performance. In \cite{cf_mrt1}, a downlink EE maximization transmit design was proposed for a Cell-Free massive MIMO system, where the beamforming vectors were designed via  maximum ratio transmission (MRT) scheme based on the uplink pilot training. In \cite{cf_mrt2}, the MRT scheme was also considered, based on which the SE for a Cell-Free massive MIMO system was maximized and analyzed. In \cite{cf_zf}, the uplink transmission SE was analyzed for a Cell-Free massive MIMO system by means of the zero-forcing (ZF) scheme and estimated channel. When integrating the ISAC and the Cell-Free massive MIMO, the pilot training becomes even more challenging in the presence of location-unknown targets which in reality can reflect the pilot signals. Although in \cite{isac_cf2} and \cite{isac_cf3}, the transmit designs were studied for the Cell-Free massive MIMO ISAC systems, they relied on the perfect channel state information assumption without uplink pilot training.

	 \vspace{-1em}
	 \subsection{Main Contributions}
	  {\it So far, how to jointly mathematically model the uplink training and downlink transmission phases in the Cell-Free massive MIMO ISAC system has never been reported in the open literature}. To fill this gap, this paper firstly presents a system model formulation for the Cell-Free massive MIMO ISAC system. It is noticeable that the propagation of wireless signals is more complicated in the Cell-Free massive MIMO ISAC system than that in traditional Cell-Free massive MIMO systems, as the wireless signals will be reflected by location-unknown targets during both uplink and downlink phases. Moreover, the C-S region defined for the Cell-Free massive MIMO ISAC  is for an in-dept trade-off study between communication and radar sensing. The main contributions of this paper are summarized as follows:
	 
	 1) A Cell-Free massive MIMO ISAC system model is proposed, where during the uplink phase, the channel between each AP and each user is estimated based on the pilot signals transmitted by the user and reflected from multiple targets, and during the downlink phase, the APs serve the users based on the estimated channel and perform target sensing based on the prior information of their locations in the meantime. In addition, the main statistics of the MIMO channel estimation error are theoretically derived in the closed-form fashion.
	 
	 2) The C-S region is defined for the Cell-Free massive MIMO ISAC system, which can be determined from the following three cases for downlink transmit beamforming design: (\rmnum{1}) the sensing-only case by minimizing the sensing beampattern matching MSE; (\rmnum{2}) the communication-only case by maximizing the sum rate; (\rmnum{3}) the ISAC case by maximizing the sum rate subject to all the achievable sensing beampattern matching MSE requirements.
	 
	 3) The transmit beamforming design problems for the three cases are respectively carried out. The sensing-only case can be directly solved as it is convex. The communication-only case is solved by a block coordinate descent (BCD)-based algorithm after the Lagrangian dual transform and  the quadratic fractional transform. The ISAC case is first reformulated similar to the communication-only case and then, solved by a successive convex approximation (SCA) based algorithm in a BCD manner.
	 
	 4) Numerical results are provided to demonstrate the C-S region, including the C-S performance trade-off characteristics and its impacts by the system parameters, and a comparison with the traditional cellular system. Moreover, the efficacy of the proposed joint C-S transmit beamforming scheme is validated together with its superior performance over the ZF scheme and the MRT scheme.

	 The rest of the paper is organized as follows. Section \Rmnum{2} formulates the system model and the associated optimization problems. Section \Rmnum{3} presents a pilot allocation scheme and a target-AP pairing scheme. Section \Rmnum{4} solves the corresponding problems to obtain the C-S region. Section \Rmnum{5} shows the numerical results. Section \Rmnum{6} concludes the paper.
	
	{\it Notations:} In this paper, $x$, ${\bf x}$, ${\bf X}$ and ${\cal X}$ are respectively denoted by scalar, vector, matrix and set. ${\rm Re}\{ \cdot  \}$ denotes the real part of a complex number or vector. $| \cdot |$   denotes the absolute value of a complex scalar. $|| \cdot ||$ denotes the two-norm for a complex vector. $(\cdot)^T$, $(\cdot)^*$ and $(\cdot)^H$ denote the transpose, conjugate and conjugate transpose, respectively. ${\mathbb C}^M$ and  ${\mathbb C}^{M \times N}$ denote the set of $M \times 1$ complex-valued vectors and $M \times N$ complex-valued matrices, respectively. ${\bf X} \succeq {\bf 0}$ denotes ${\bf X}$ is a positive semidefinite matrix. ${\bf a} \sim \mathcal{CN}({\bm \mu}, {\bm \Sigma})$ denotes that ${\bf a}$ is a complex valued circularly symmetric Gaussian random variable with mean  ${\bm \mu}$ and covariance matrix ${\bm \Sigma}$. $\mathbb{E}\{ \cdot \}$ denotes the expectation  for a random variable. $ \{  {\bf x}_{ij}  \}$ and $\{ {\bf x}_{ij} \}_i$ denote all admissible $\{ i,j\}$ for ${\bf x}_{ij}$ and all admissible $i$ for ${\bf x}_{ij}$ with fixed $j$, respectively. 
	
	 \vspace{-1em}
	\section{System Model}
	\begin{figure}
	\setlength{\abovecaptionskip}{0.cm}
		\begin{center}
			\centerline{\includegraphics[width=0.9\textwidth]{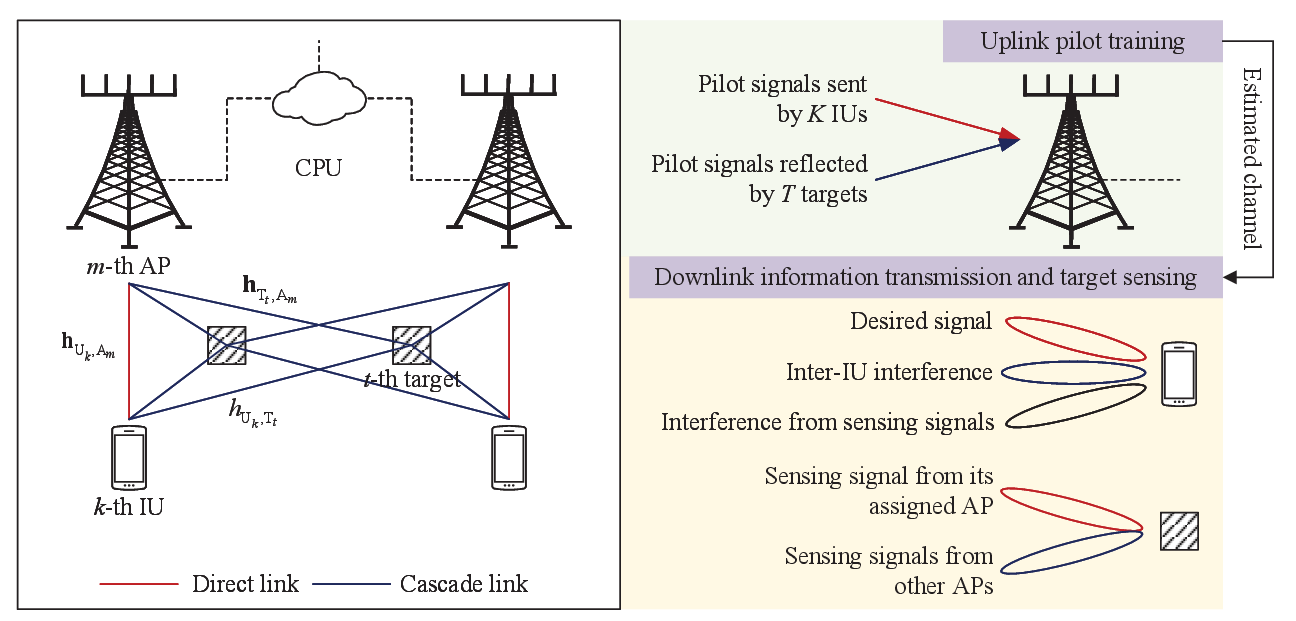}}
			\caption{ A Cell-Free Massive MIMO ISAC system.}
			\label{system}
		\end{center}
		\vspace{-1.7cm}
	\end{figure}
	Consider a Cell-Free Massive MIMO ISAC system as shown in Figure \ref{system}, where a CPU coordinates $M$ $N_{\rm T}$-antenna APs to jointly serve $K$ single-antenna information users (IUs) and detect $T$ point-like device-free targets over the same time and frequency resource. The considered system operates in a time-division duplex mode and each coherence interval is enough to cover $\tau_{\rm c}$ symbols, constituted by $\tau_{\rm p}$ symbols for  uplink pilot training and $( \tau_{\rm c}-\tau_{\rm p} )$ symbols for downlink information transmission and target sensing. In order to achieve high-resolution sensing performance, each target is assigned to one AP to sense, while each AP can sense at most one target. Besides, the signals transmitted by IUs and APs can also be reflected by targets during both uplink and downlink phases. For clarity, let $\mathcal{M} \triangleq \{1,2,...,M \}$, $\mathcal{K} \triangleq \{1,2,...,K \}  $ and $\mathcal{T} \triangleq \{1,2,...,T \}  $  denote the sets of APs, IUs and targets, respectively. 
	
	 \vspace{-1em}
	\subsection {\it Channel Model}
	
	The quasi-static block fading model is adopted, that is, the channels are static and frequency flat in each coherence interval. The distance-dependent path-loss function denoted by $\beta(\cdot )$ can be expressed as \cite{bc_massive_mimo1}
	\begin{flalign}
		\beta(d) = 10^{-(128.1 + 37.6 \log_{10} (d) ) /20}  ,
	\end{flalign}
	where $d$ denotes the distance between the transceivers. The channel between the $m$-th AP and the $k$-th IU is given by
	\begin{flalign}
		{\bf h}_{{\rm U}_k, {\rm A}_m} = { \beta \left(  d_{{\rm U}_k, {\rm A}_m    }   \right)    } {\bf q}_{ {\rm U}_k , {\rm A}_m },
	\end{flalign} 
	where $d_{{\rm U}_k, {\rm A}_m} \triangleq \sqrt{(x_{ {\rm U}_k} - x_{{\rm A}_m})^2 + (y_{{\rm U}_k}-y_{{\rm A}_m})^2     } $ denotes the distance between the $k$-th IU and the $m$-th AP with $(x_{{\rm U}_k}, y_{{\rm U}_k} )$ and $( x_{{\rm A}_m}, y_{{\rm A}_m})$ denoting the coordinates of the $k$-th IU and the $m$-th AP, respectively; ${\bf q}_{ {\rm U}_k, {\rm A}_m} \in \mathbb{C}^{N_{\rm T}}$ represents small-scale  Rayleigh fading, i.e., ${\bf q}_{ {\rm U}_k, {\rm A}_m}  \sim \mathcal{CN}({\bf 0}, {\bf I}_{N_{\rm T} } )$.
	
	In practice, the coordinates of the targets may not be exactly known. Based on the pre-detected technology \cite{pre_detect}, the APs can have prior knowledge of the $t$-th target's coordinate, denoted by $(\bar{x}_{{\rm T}_t}, {\bar y}_{{\rm T}_t})$. The channel between the $m$-th AP and the $t$-th target is modeled as
	\begin{flalign}
		{\bf h}_{ {\rm T}_t, {\rm A}_m  } = {\beta({d_{{\rm T}_t, {\rm A}_m  }})} {\bf q}_{{\rm T}_t, {\rm A}_m }, 
	\end{flalign}
	where $d_{{\rm T}_t, {\rm A}_m}$ denotes the distance between the $t$-th target and the $m$-th AP, and it is assumed that $\beta(d_{{\rm T}_t, {\rm A}_m}) \sim \mathcal{N}(\beta(\bar{d}_{{\rm T}_t,{\rm A}_m}), \sigma_{ {\rm T}_t, {\rm A}_m}^2) $ in which $\bar{d}_{{\rm T}_t, {\rm A}_m} \triangleq \sqrt{ (\bar{x}_{{\rm T}_t} - x_{{\rm A}_m}  )^2 + ( \bar{y}_{{\rm T}_t} - y_{{\rm A}_m} )^2 }$; ${\bf q}_{{\rm T}_t, {\rm A}_m} \sim \mathcal{CN}(  \bar{ {\bf q}}_{{\rm T}_t, {\rm A}_m }, \chi_{{\rm T}_t, {\rm A}_m}^2 {\bf I}_{N_{\rm T}  }  )$ denotes the steering vector between the $t$-th target and the $m$-th AP provided that the sensing channel is line of sight \cite{los}, where $\bar{\bf q}_{ {\rm T}_t, {\rm A}_m }$ is given by 
	\begin{flalign*}
		\bar{\bf q}_{{\rm T}_t, {\rm A}_m} = (1, e^{j 2 \pi \frac{d}{\lambda} \sin(\theta_{tm})  },..., e^{j 2 \pi \frac{d}{\lambda}(N_{\rm T}-1) \sin(\theta_{tm})  }   ) \in \mathbb{C}^{{N}_{\rm T}},
	\end{flalign*}
	in which $d$ and $\lambda$ denote the spacing between two adjacent antennas and the wavelength of the signal, and $\theta_{tm}$ denotes the direction of arrival (DOA) from the $m$-th AP to the $t$-th target, where
	\begin{flalign}
		\theta_{tm} = \arccos \frac{y_{{\rm A}_m} -\bar{y}_{{\rm T}_t}  }{\sqrt{ (x_{{\rm A}_m}- \bar{x}_{{\rm T}_t }   )^2 + (y_{{\rm A}_m} -\bar{y}_{{\rm T}_t})^2    }   }. \label{DOA}
	\end{flalign}

	As previously mentioned, the targets may reflect the signals from/to IUs. The channel between the $t$-th target and the $k$-th IU is given by 
	\begin{flalign}
		{h}_{{\rm U}_k, {\rm T}_t   } = {\beta(d_{{\rm U}_k, {\rm T}_t   })} {q}_{ {\rm U}_k, {\rm T}_t    },
	\end{flalign}
	where $d_{{\rm U}_k, {\rm T}_t }$ denotes the distance between the $t$-th target and the $k$-th IU;  ${q}_{{\rm U}_k, {\rm T}_t  } \sim \mathcal{CN}(0, 1)$ represents Rayleigh fading between the $t$-th target and the $k$-th IU. Similarly, it is assumed $\beta(d_{{\rm U}_k,{\rm T}_t }) \sim \mathcal{N}( {\beta}(\bar{d}_{{\rm U}_k,{\rm T}_t }), \sigma_{{\rm U}_k, {\rm T}_t  }^2) $, where $\bar{d}_{{\rm U}_k, {\rm T}_t} \triangleq \sqrt{(x_{{\rm U}_k} - \bar{x}_{ {\rm T}_t  })^2  + (y_{{\rm U}_k} - \bar{y}_{{\rm T}_t}  )^2 } $. 
	
	 \vspace{-1em}
	\subsection{\it Uplink Pilot Training}
	
	In the uplink pilot training phase, the $k$-th IU transmits a pilot sequence of $\tau_{\rm p}$ symbols to
	APs to facilitate the channel estimation. To enhance the utilization of the pilot sequence, it is assumed that ${\tau_{\rm p}} < K$, indicating that the pilot sequence must be reused. Denote $\mathcal{P}_k$ by the set of indices of the IUs that share the same pilot sequence with the $k$-th IU. Note that if the $k$-th IU shares the same pilot sequence with the $k'$-th IU, it holds that  $\mathcal{P}_k =  \mathcal{P}_{k'}$. Let $\sqrt{\tau_{\rm p}} {\bm \phi}_k \in \mathbb{C}^{\tau_{\rm p}}$ with $\Vert {\bm \phi}_k    \Vert^2  = 1$ denote the pilot sequence allocated to the $k$-th IU, with the following property:
	\begin{flalign*}
			{\bm \phi}_{k'}^H {\bm \phi}_k = 
		\left\{ \begin{array}{l}
			1,~   {\rm if}~ k' \in \mathcal{P}_k,\\
			0,~   {\rm if}~ k' \notin \mathcal{P}_k.
		\end{array} \right.
	\end{flalign*}
	Then, the received signal at the $m$-th AP can be expressed as
	\begin{flalign}
		{\bf Y}_{{\rm p}, m} = &\sum_{k \in \mathcal{K} }  \sqrt{p_{\rm p} \tau_{\rm p}} \left( {\bf h}_{{\rm U}_k, {\rm A}_{m} } + \sum_{t \in \mathcal{T} } \alpha_t h_{ {\rm U}_k, {\rm T}_t}  {\bf h}_{{\rm T}_t, {\rm A}_m  }   \right)  {\bm \phi}_k^H  
		+ {\bf N}_{{\rm p}, m},  \label{receive1}
	\end{flalign}
	where $p_{\rm p}$ denotes the normalized SNR of each pilot symbol; $\alpha_t$ denotes the reflection coefficient of the $t$-th target since it can reflect the signals transmitted by IUs; ${\bf N}_{{\rm p}, m} \in \mathbb{C}^{N_{\rm T} \times {\tau_{\rm p}}   } $ denotes the normalized additive white Gaussian noise (AWGN) at the $m$-th AP and all the elements of ${\bf N}_{{\rm p}, m}$ are independent identically distributed (i.i.d.) with the distribution $\mathcal{CN}(0, 1)$. 
	
	To estimate the channel between the $m$-th AP and the $k$-th IU, the received signal in (\ref{receive1}) can be projected on ${\bm \phi}_k$ as follows
	\begin{flalign}
		{\bf y}_{ {\rm p}, mk} \triangleq {\bf Y}_{{\rm p}, m} {\bm \phi}_k  = & \sqrt{p_{\rm p}  \tau_{\rm p}  } {\bf h}_{mk} + \sum_{k' \in \mathcal{P}_k \setminus \{ k  \}  } \sqrt{p_{\rm p}  \tau_{\rm p}} {\bf h}_{mk'}   
		 + {\bf n}_{{\rm p}, mk}, \label{y}
	\end{flalign}
	where ${\bf h}_{mk} \triangleq {\bf h}_{{\rm U}_k, {\rm A}_{m} } + \sum_{t \in \mathcal{T} } \alpha_t h_{ {\rm U}_k, {\rm T}_t}  {\bf h}_{{\rm T}_t, {\rm A}_m  }$ denotes the aggregated channel between the $m$-th AP and $k$-th IU which includes one direct link and $T$ cascade links; ${\bf n}_{{\rm p}, mk } \triangleq  {\bf N}_{ {\rm p}, m}{\bm \phi}_k \sim \mathcal{CN}({\bf 0}, {\bf I}_{N_{\rm T}})  $ denotes the projection of ${\bf N}_{{\rm p}, m}$ on ${\bm \phi}_k$. Instead of  estimating ${\bf h}_{{\rm U}_k, {\rm A}_{m} }, h_{ {\rm U}_k, {\rm T}_t}$ and ${\bf h}_{{\rm T}_t, {\rm A}_m  }$ separately, we estimate the aggregated channel ${\bf h}_{mk}$, which is more suitable for practical implement without affecting the downlink transmission. To facilitate channel estimation, the cross-correlation matrix of any two aggregated channels is given in the following proposition. 
	\begin{mypro}
		Let $\{ {\bf h}_{mk}\}$ be the aggregated channel defined in (\ref{y}). Then
		\begin{flalign*}
			\mathbb{E}  \left\{  {\bf h}_{mk} {\bf h}_{m'k'}^H  \right\} = 
			\left\{ \begin{array}{l}
				{\bm \Phi}_{mk} ,~ {\rm if} ~~m' = m, k' = k,\\
				{\bm \Phi}_{mm'k} ,~ {\rm if} ~~m' \neq m, k' = k, \\
				{\bf 0}, ~~~ {\rm otherwise},  \\
			\end{array} \right.
		\end{flalign*}
		where ${\bm \Phi}_{mk}$ and ${\bm \Phi}_{mm'k}$ are given as follows, respectively,
			\begin{flalign}
				{\bm \Phi}_{mk} &= \beta^2 \left( d_{{\rm U}_k, {\rm A}_m} \right) {\bf I}_{N_{\rm T}  } + \sum_{t \in \mathcal{T}}    \left(      \alpha_t^2 \left( \beta^2\left({ \bar{d}_{{\rm U}_k, {\rm T}_t   }   }\right) + \sigma_{{\rm U}_k, {\rm T}_t  }^2  \right)  \left( {\beta^2({  \bar{d}_{{\rm T}_t, {\rm A}_m  }})} + \sigma_{ {\rm T}_t, {\rm A}_m}^2 \right)  \right.  \nonumber \\
				&\left.~~  \times \left(  \bar{\bf q}_{{\rm T}_t, {\rm A}_m }  \bar{\bf q}_{{\rm T}_t, {\rm A}_m }^H + \chi_{{\rm T}_t, {\rm A}_m}^2 {\bf I}_{ N_{\rm T} }  \right)  \label{phi_mk} \right),
				  \\
				{\bm \Phi}_{mm'k} &= \sum_{t \in \mathcal{T}} \alpha_t^2 \left( \beta^2\left({ \bar{d}_{{\rm U}_k, {\rm T}_t   }   }\right) + \sigma_{{\rm U}_k, {\rm T}_t  }^2  \right)   {\beta({  \bar{d}_{{\rm T}_t, {\rm A}_m  }})}  {\beta({  \bar{d}_{{\rm T}_t, {\rm A}_{m'}  }})}   \bar{\bf q}_{{\rm T}_t, {\rm A}_m }  \bar{\bf q}_{{\rm T}_t, {\rm A}_{m'} }^H   \label{phi_mm'k } .
			\end{flalign}	
	\end{mypro}
	\begin{IEEEproof}
		The proof is given in Appendix A.
	\end{IEEEproof}
	
	By applying the linear minimum MSE method \cite{lmmse}, the estimation of the aggregated channel between the $k$-th IU and $m$-th AP, denoted by $\widehat{\bf h}_{mk}$, can be expressed as
	\begin{flalign}
		\widehat{\bf h}_{mk} &= \displaystyle\underbrace{\mathbb{E}\left\{ {\bf h}_{mk} {\bf y}_{{\rm p},mk}^H   \right\}\mathbb{E}^{-1}\left\{ {\bf y}_{{\rm p},mk} {\bf y}_{{\rm p},mk}^H  \right\}}_{ \triangleq {\bf C}_{mk}} {\bf y}_{{\rm p},mk} .  \label{hat_h}
	\end{flalign}
	
	\begin{mypro}
		The mean and auto-correlation matrix of the channel estimations $\{ \widehat{\bf h}_{mk}  \}$ defined in (\ref{hat_h})  are given by
		\begin{flalign}
			\mathbb{E}\left\{ \widehat{\bf h}_{mk}   \right\}  = {\bf 0}~ {\rm and}~\mathbb{E}\left\{   \widehat{\bf h}_{mk}  \widehat{\bf h}_{mk}^H \right\}  =  \sqrt{p_{\rm p} \tau_{\rm p}} {\bf C}_{mk} {\bm \Phi}_{mk},  \nonumber
		\end{flalign}
		respectively, and ${\bf C}_{mk}$ defined in (\ref{hat_h}) can be expressed in a closed form as follows
		\begin{flalign}
			{\bf C}_{mk} = \sqrt{p_{\rm p}  \tau_{\rm p}} {\bm \Phi}_{mk} \left( p_{\rm p}  \tau_{\rm p} \sum_{k' \in \mathcal{P}_k  }{\bm \Phi}_{mk'} + {\bf I}_{N_{\rm T} }  \right)^{-1}. \nonumber
		\end{flalign}
	\end{mypro}
	\begin{IEEEproof}
		The proof is given in Appendix B.
	\end{IEEEproof}
	
	Define the channel estimation error as
	\begin{flalign}
		{\bf e}_{mk} \triangleq {\bf h}_{mk} - \widehat{\bf h}_{mk}, \label{e}
	\end{flalign}
	and its mean and second-order statistics are analyzed in the following proposition.
	
	\begin{mypro}
		The mean and cross-correlation matrix of the channel estimation error $\{ {\bf e}_{mk} \}$ defined in (\ref{e}) are given as follows
	\begin{flalign*}
			{	\mathbb{E}\left\{  {\bf e}_{mk}  \right\}	= {\bf 0},~{\rm and}~ \mathbb{E}  \left\{  {\bf e}_{mk} {\bf e}_{m'k}^H  \right\} = } 
		\left\{ \begin{array}{l}
				{\bm \Theta}_{mk} ,~ {\rm if} ~~m' = m,\\
				{\bm \Theta}_{mm'k},~ {\rm if} ~~m' \neq m,
		\end{array} \right.
	\end{flalign*}
		where ${\bm \Theta}_{mk} \triangleq {\bm \Phi}_{mk} -  \sqrt{p_{\rm p} \tau_{\rm p}} {\bf C}_{mk} {\bm \Phi}_{mk} $ and ${\bf \Theta}_{mm'k}$ is given by
			\begin{flalign}
				{\bf \Theta}_{mm'k} \triangleq  {\bf \Phi}_{mm'k} - \sqrt{p_{\rm p} \tau_{\rm p}  } {\bf \Phi}_{mm'k} {\bf C}_{m'k}^H- \sqrt{p_{\rm p}\tau_{\rm p}} {\bf C}_{mk} {\bf \Phi}_{mm'k} + p_{\rm p} \tau_{\rm p} {\bf C}_{mk} \sum_{k' \in \mathcal{P}_k  } {\bf \Phi}_{mm'k'}
				{\bf C}_{m'k}^H. \label{channel_error}
			\end{flalign}
	\end{mypro}
	\begin{IEEEproof}
		The proof is given in Appendix C.
	\end{IEEEproof}

         Once one realization of ${\bf y}_{{\rm p},mk}$ is received at the $m$-th AP, the corresponding channel estimation $\widehat{\bf h}_{mk}$ can be obtained via (\ref{hat_h}) and Proposition 2. Next, let us use the obtained $\{ \widehat{\bf h}_{mk}\}$ and the first-order and second-order statistics of $\{ {\bf e}_{mk} \}$ provided in Proposition 3 to investigate the performance of downlink information transmission and target sensing.
	

	 \vspace{-1em}
	\subsection{\it Downlink Information Transmission and Target Sensing    }
	
	In the downlink phase, the APs operate in the dual-function mode, i.e., information transmission and target sensing. It is assumed that $M \geq  T$ in the Cell-Free system. Therefore, there may exist some APs only performing the information transmission. Let $\mathcal{M}_{\rm T}$ ($\mathcal{M}_{\rm I}$) denotes the set of APs with (without) the target sensing task, namely, $\mathcal{M}_{\rm T} \cup \mathcal{M}_{\rm I} = \mathcal{M}$ and $\mathcal{M}_{\rm T} \cap \mathcal{M}_{\rm I} = \varnothing$.
	
	For the $m$-th AP in $\mathcal{M}_{\rm I}$, the transmit signal is given by
	\begin{flalign}
		{\bf x}_m  = \sqrt{p_{m}} \sum_{k \in \mathcal{K} } {\bf w}_{mk} s_k,~ \forall m \in \mathcal{M}_{\rm I}, \label{trans_without_target}
	\end{flalign}
	where $p_{m}$ denotes the normalized signal power in the downlink phase; $s_k \in \mathbb{C}$ denotes the data symbol intended for the $k$-th IU with $\mathbb{E}\{  |s_k|^2  \} = 1$; ${\bf w}_{mk}\in \mathbb{C}^{N_{\rm T} }$  denotes the corresponding beamforming vector. The normalized power budget of the $m$-th AP is given by $\mathbb{E}\left\{  ||   {\bf x}_m ||^2   \right\} \leq p_{m} $, and thus, it holds that
	\begin{flalign}
		\sum_{k \in \mathcal{K} } {\bf w}_{mk}^H {\bf w}_{mk} \leq 1, ~ \forall m \in \mathcal{M}_{\rm I}.  \label{power1}
	\end{flalign}

	For the $m$-th AP in $\mathcal{M}_{\rm T}$, the transmit signal is given by
	\begin{flalign}
		{\bf x}_m =  \sqrt{p_{m}} \sum_{k \in \mathcal{K} }  {\bf w}_{mk} s_k + \sqrt{p_{m}} {\bf z}_m,~ \forall m \in \mathcal{M}_{\rm T}, \label{ trans_with_target }
	\end{flalign}
	where ${\bf z}_m \in \mathbb{C}^{N_{\rm T}}$ is the dedicated random sensing signal transmitted by the $m$-th AP, which is statistically independent of the data symbols $\{  s_k \}$ with $\mathbb{E}\{ {\bf z}_m {\bf z}_m^H  \} = {\bf Z}_m$. Note that the AP generates sensing signal in potential directions of targets. However, only the estimated locations of the targets are available, and the exact distance between each target and its associated AP is unknown. Therefore, in order to achieve the best sensing performance, each AP has to use its utmost power, to enlarge its probing range. Thus, we have
	\begin{flalign}
		& \mathbb{E}\{  || {\bf x}_m ||^2  \} = p_{m}  
		 \Rightarrow \sum_{k \in \mathcal{K} } {\bf w}_{mk}^H {\bf w}_{mk} + {\rm Tr}\{ {\bf Z}_m \} = 1, \forall m \in \mathcal{M}_{\rm T}. \label{power2}
	\end{flalign}

	The received signal at the $k$-th IU is the superposition of the signals transmitted by all the APs, which,  according to (\ref{trans_without_target}) and (\ref{ trans_with_target }), can be expressed as
	\begin{flalign}
		&y_{{\rm d}, k} = \sum_{m \in \mathcal{M}} {\bf h}_{mk}^H {\bf x}_m + n_{{\rm d},k} \nonumber \\
		&~= \sum_{m \in \mathcal{M} } \sqrt{p_{m}} \widehat{\bf h}_{mk}^H {\bf w}_{mk} s_k +  \sum_{m \in \mathcal{M}} \sqrt{p_{m}} {\bf e}_{mk}^H {\bf w}_{mk} s_k  
		 + \sum_{k' \neq k}  \sum_{m \in \mathcal{M}}\sqrt{p_{m}} \left(  \widehat{\bf h}_{mk}^H {\bf w}_{mk'} s_{k'} + {\bf e}_{mk}^H {\bf w}_{mk'} s_{k'}   \right) 
		 \nonumber \\
		&~~~~~   +  \sum_{m \in \mathcal{M}_{\rm T} }\sqrt{p_{m}} \left( \widehat{\bf h}_{mk}^H {\bf z}_{m} + {\bf e}_{mk}^H {\bf z}_m    \right)  + n_{{\rm d}, k},  \label{receive}
	\end{flalign}
	where $n_{{\rm d},k} \in \mathcal{CN}(0,1)$ denotes the normalized AWGN at the $k$-th IU. Due to the presence of random channel errors $\{ {\bf e}_{mk} \}$ in the received signal $y_{{\rm d}, k}$ (cf. (\ref{receive})), the design of $\{ {\bf w}_{mk}, {\bf Z}_m \}$ is almost prohibitive. Next, let us present how we proceed with this design.
	
	Similar to \cite{js}, a computationally tractable lower bound on the IU's ergodic rate can be derived via Jensen's Inequality over ${\bf e}_{mk}$. That is,  $\mathbb{E} \{ \log(1+{\rm SINR})   \} \geq \log( 1 + 1/\mathbb{E}\{ 1/{\rm SINR}  \}  )$. Denote  $r_k\left( \left\{  {\bf w}_{mk}, {\bf Z}_m   \right\}  \right)$ by the lower bound on the achievable $k$-th IU's ergodic rate, which is given by
	\begin{flalign}
		r_k \left(  \left\{   {\bf w}_{mk}, {\bf Z}_m  \right\} \right) = \log_2 \left( 1+ \frac{\left|\sum_{m \in \mathcal{M}} \sqrt{p_m} \widehat{\bf h}_{mk}^H {\bf w}_{mk} \right|^2 }  { A_k\left( \left\{   {\bf w}_{mk}, {\bf Z}_m    \right\}   \right)   }  \right),   \nonumber
	\end{flalign}
	where $A_k\left( \left\{   {\bf w}_{mk}, {\bf Z}_m    \right\}   \right)$ is given in (\ref{noise}), composed of the inter-IU interference, other interference terms induced from the channel estimation errors and/or sensing signals, and AWGN.
	\begin{figure*}
		\begin{flalign}
			&A_k\left( \left\{   {\bf w}_{mk}, {\bf Z}_m    \right\}   \right)  \triangleq \nonumber \\
			& \mathbb{E} \left\{ \left| \sum_{m \in \mathcal{M}} \sqrt{p_{m}} {\bf e}_{mk}^H {\bf w}_{mk} \right|^2 \right\} + \sum_{k' \neq k} \left| \sum_{m \in \mathcal{M}}   \sqrt{ p_m } \widehat{\bf h}_{mk}^H {\bf w}_{mk'}   \right|^2 + \sum_{k' \neq k} \mathbb{E}\left\{ \left|  \sum_{m \in \mathcal{M}} \sqrt{p_m} {\bf e}_{mk}^H {\bf w}_{mk'}  \right|^2  \right\}  \nonumber \\
			&~~~~~~~~
			+  \sum_{m \in \mathcal{M}_{\rm T} } p_m \widehat{\bf h}_{mk}^H {\bf Z}_m \widehat{\bf h}_{mk}	  + \sum_{m \in \mathcal{M}_{\rm T} }  p_m \mathbb{E} \left\{  {\bf e}_{mk}^H {\bf Z}_m {\bf e}_{mk}   \right\}  +1  \nonumber \\
			& = \sum_{m \in \mathcal{M}} p_m {\bf w} _{mk}^H {\bf \Theta}_{mk} {\bf w}_{mk} + \sum_{m \in \mathcal{M}} \sum_{m' \neq m} \sqrt{p_m p_{m'}  } {\bf w}_{mk}^H {\bf \Theta}_{mm'k} {\bf w}_{m'k} + \sum_{k' \neq k} \left| \sum_{m \in \mathcal{M}}   \sqrt{ p_m } \widehat{\bf h}_{mk}^H {\bf w}_{mk'}   \right|^2   \nonumber \\
			&~~~~~~~~
			+ \sum_{k' \neq k} \sum_{m \in \mathcal{M}} p_m {\bf w}_{mk'}^H {\bf \Theta}_{mk} {\bf w}_{mk'} + \sum_{k' \neq k} \sum_{m \in \mathcal{M}} \sum_{m' \neq m} \sqrt{p_m p_{m'}} {\bf w}_{mk'}^H {\bf \Theta}_{mm'k} {\bf w}_{m'k'} \nonumber  \\
			&~~~~~~~~
			+ \sum_{m \in \mathcal{M}_{\rm T} } p_m \widehat{\bf h}_{mk}^H {\bf Z}_m \widehat{\bf h}_{mk}	  + \sum_{m \in \mathcal{M}_{\rm T} }  p_m {\rm Tr} \left\{  {\bf \Theta}_{mk} {\bf Z}_m  \right\}  +1
			\label{noise}  
		\end{flalign}
		\hrule
	\end{figure*}
	
	Denote the set of targets assigned to the $m$-th AP  by $\mathcal{T}_m$, for which $\mathcal{T}_m = \varnothing$ for all $m \in \mathcal{M}_{\rm I}$ and $ {\mathcal{T}_{m}}$ is non-empty for all $ m \in \mathcal{M}_{\rm T}$. Empirically, the best sensing performance of an AP can be obtained when the target is at the direction with maximum gain of its antennas beampattern \cite{s1}. Based on such an observation, one AP is assigned to sense at most one target as the number of APs is usually larger than that of the targets. That is, $| \mathcal{T}_m| = 1$, $\forall m \in \mathcal{M}_{\rm T}$. Provided that the $t$-th target is the element of $\mathcal{T}_m$, the $m$-th AP senses the $t$-th target based on the priori knowledge of the DOA (also known as the DOA estimation), i.e., $\theta_{tm}$, which can be obtained by (\ref{DOA}). Particularly, $\theta_{tm}$ is utilized to guide the generation of the $m$-th AP's beampattern. For a direction $\theta \in [-\pi/2, \pi/2]$ of an AP, the steering vector can be expressed as ${\bf a}(\theta) = [1,e^{j 2\pi \frac{d}{\lambda} \sin(\theta) },..., e^{j 2\pi (N_{\rm T}-1 ) \frac{d}{\lambda} \sin(\theta) }]^T$. Then, the amplitude of the $m$-th AP's beampattern in the direction $\theta$ can be expressed as \cite{s1}
	\begin{flalign}
		P_m(\theta)  &= \mathbb{E} \left\{   \left|   {\bf a}^H(\theta)  {\bf x}_m  \right|^2   \right\} 
		= p_m \sum_{k \in \mathcal{K} } \left| {\bf a}^H(\theta) {\bf w}_{mk}  \right|^2 + p_m {\bf a}^H(\theta) {\bf Z}_m {\bf a}(\theta).
	\end{flalign} 
    Besides, with $\theta_{tm}$, an ideal beampattern is given by \cite{s2}, 
	\begin{numcases}
		{ \widetilde{P}_m (\theta) = }
		1,~\left|  \theta - \theta_{ t m}   \right| \leq \frac{ \Delta \theta }{2},         \nonumber \\
		0,~{\rm otherwise},             \nonumber
	\end{numcases}
	where $\Delta \theta$ denotes the width of the ideal beampattern.
	
    A practical beampattern design is adopted for each AP in $\mathcal{M}_{\rm T}$ as follows. Let $\{\bar{\theta}_n \}_{n=1}^{N}$ be a considered set of $N$ sampled directions over $[-\pi/2, \pi/2]$. The desirable beampattern can be obtained by minimizing the MSE between the generated beampattern and the ideal beampattern, which is denoted by $\mathcal{E}_m \left(  \eta_m, \{ {\bf w}_{mk} \}, {\bf Z}_m \right) $ and given by (\ref{mse}), where $\mathcal{N} \triangleq \{1,2,...,N\}$ denotes the index set of the sampled directions, and $\eta_m$ denotes the scaling factor.
	\begin{figure*}
		\begin{flalign}
			\mathcal{E}_m \left(  \eta_m, \{ {\bf w}_{mk} \}, {\bf Z}_m \right) =  \frac{1}{N} \sum_{n \in \mathcal{N} } \left|\eta_m   \widetilde{P}_m (\bar{\theta}_n) - p_m {\bf a}^H(\bar{\theta}_n) \left( \sum_{k \in \mathcal{K} } {\bf w}_{mk} {\bf w}_{mk}^H + {\bf Z}_m   \right)  {\bf a}( \bar{\theta}_n )      \right|^2 \label{mse}
		\end{flalign}
		\hrule
	\end{figure*}

	 \vspace{-1em}
	\subsection {\it Communication-sensing Region}
	
\begin{figure}
		\setlength{\abovecaptionskip}{0.cm}
		\begin{center}
			\centerline{\includegraphics[width=.5\textwidth]{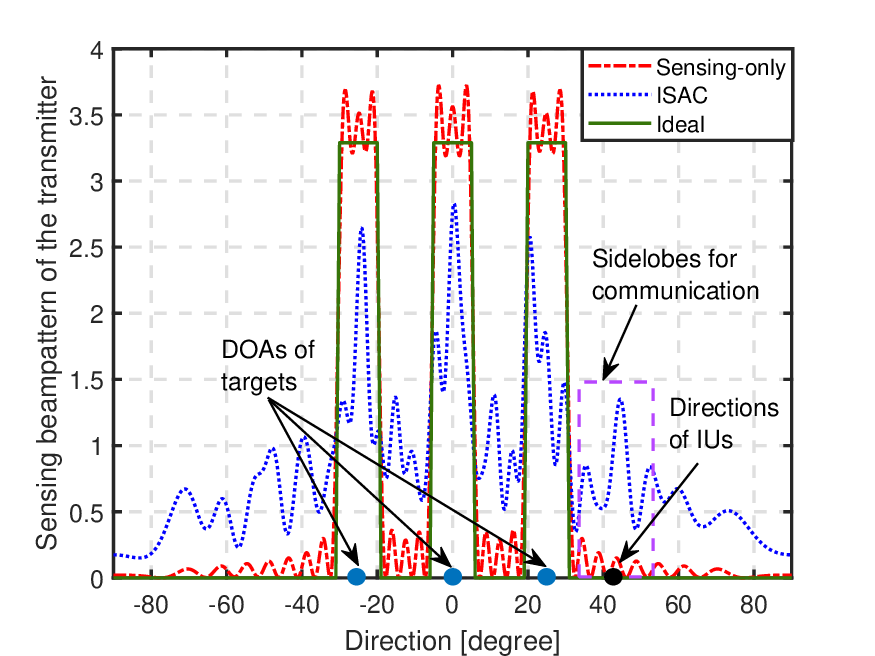}}
			\caption{ Illustration of the trade-off between communication and sensing.}
			\label{beam}
		\end{center}
		\vspace{-2cm}
	\end{figure}

	A special characteristic for this case of joint communication and sensing is the trade-off between them. An example is illustrated in Figure \ref{beam}, from which one can observe that more power in sidelobes for communication results in smaller probing range of sensing. This motivates us to investigate the {\emph{C-S region}} to explore the optimal transmit design for the ISAC system. The C-S region is defined as
	\begin{flalign}
		{\mathcal {R}}_{\text{C-S}   }\left( \{ {\bm \phi}_k  \}, \{ \mathcal{T}_m  \}   \right)  \triangleq 
		&\Big\{ (S, R): S \geq  \frac{1}  {M_{\rm T}} \sum_{m \in \mathcal{M}_{\rm T}} \mathcal{E}_m \left(  \eta_m, \{ {\bf w}_{mk} \}, {\bf Z}_m \right),  R \leq   \sum_{k \in \mathcal{K}} r_k \left(  \left\{   {\bf w}_{mk}, {\bf Z}_m  \right\} \right) ,  \nonumber \\
		& \left\{  {\bf Z}_m \right\} \succeq  {\bf 0} , {\rm (\ref{power1}), (\ref{power2}) }
		    \Big\},  \label{c-s}
	\end{flalign} 
	where $M_{\rm T}=|{\cal M}_{\rm T}|$ denotes the number of APs required for target sensing.

	The C-S region is achieved by solving the transmit design problems for the following three cases, i.e., the sensing-only case, the communication-only case and the ISAC case.
	
	\noindent 1) {\it The sensing-only case}: This case without any IUs actually corresponds to the special case of $\{  {\bf w}_{mk} \}=\{ {\bf 0} \}$. The Cell-Free system aims to obtain the optimal sensing performance ${S_{\min}}$ by solving the following problem:
	\begin{subequations}
		\begin{align}
			{\bf P_1:}~S_{\rm min} \triangleq &\mathop {\min }\limits_{\left\{ \eta_m, {\bf Z}_m      \right\}}
			\sum_{m \in \mathcal{M}_{\rm T} } \frac{\mathcal{E}_m \left(  \eta_m, \{ {\bf 0} \}, {\bf Z}_m \right)}{M_{\rm T}} \\
			& {\rm s.t.}~ {\rm Tr}\{ {\bf Z}_m  \} =1, ~\forall m \in \mathcal{M}_{\rm T},  \\
			&~~~~~  {\bf Z}_m  \succeq {\bf 0} , ~\forall m \in \mathcal{M}_{\rm T}.
		\end{align}
	\end{subequations}
    
	\noindent 2) {\it The communication-only case}: This case without any targets actually corresponds to the special case of $\{ {\bf Z}_m\} = \{ {\bf 0} \}$. The Cell-Free system aims to obtain the optimal communication performance ${R_{\max}}$ by solving the following problem:
	\begin{subequations}
		\begin{align}
			{\bf P_2:}~{R_{\max}} \triangleq &\mathop {\max }\limits_{\left\{ {\bf w}_{mk}    \right\}}
			\sum_{k \in \mathcal{K}} r_k \left(  \left\{   {\bf w}_{mk}, {\bf 0}  \right\} \right) \\
			&{\rm s.t.}~ \sum_{k \in \mathcal{K}} {\bf w}_{mk}^H {\bf w}_{mk} \leq 1, ~\forall m \in \mathcal{M}.
		\end{align}
	\end{subequations}
	
	\noindent 3) {\it The ISAC case}: Apart from the previous two cases, this case with IUs and targets co-existing needs to determinate the coupling boundary of the C-S region within $R \in (0, R_{\max})$ and $S \in (S_{\min}, \infty)$ by solving the following problem:
	\begin{subequations}
		\begin{align}
			{\bf P_3:} &\mathop {\max }\limits_{\left\{ \eta_m, {\bf w}_{mk}, {\bf Z}_m    \right\}}   \label{p_3_a}
			\sum_{k \in \mathcal{K}} r_k \left(  \left\{   {\bf w}_{mk}, {\bf Z}_m  \right\} \right) \\
			&{\rm s.t.}~ \sum_{m \in \mathcal{M}_{\rm T}}  \frac{ \mathcal{E}_m \left(  \eta_m, \{ {\bf w}_{mk} \}, {\bf Z}_m \right)}  { M_{\rm T}  } \leq \delta ,  \label{p_3_b}   \\ 
			&~~~~~  {\bf Z}_m  \succeq {\bf 0} , ~\forall m \in \mathcal{M}_{\rm T},  \label{p_3_c}  \\
			&~~~~~  {\rm (\ref{power1}), (\ref{power2}) }, \nonumber  
		\end{align}
	\end{subequations}
	where $\delta$ denotes the sensing beampattern matching MSE requirement.
	
	\begin{remark}
		Alternatively, the C-S region can be obtained by solving the following problem instead of Problem ${\bf P_3}$
		\begin{subequations}
			\begin{align}
				{\bf P_4:} &\mathop {\min }\limits_{\left\{ \eta_m, {\bf w}_{mk}, {\bf Z}_m    \right\}}
				\sum_{m \in \mathcal{M}_{\rm T}}  \frac{ \mathcal{E}_m \left(  \eta_m, \{ {\bf w}_{mk} \}, {\bf Z}_m \right)}  { M_{\rm T}  }
				\\ 
				&{\rm s.t.}~ \sum_{k \in \mathcal{K}} r_k \left(  \left\{   {\bf w}_{mk}, {\bf Z}_m  \right\} \right) \geq \zeta,   \\
				&~~~~~   {\bf Z}_m  \succeq {\bf 0} , ~\forall m \in \mathcal{M}_{\rm T},\\
				&~~~~~  {\rm (\ref{power1}), (\ref{power2}) }, \nonumber  
			\end{align}
		\end{subequations}
		where $\zeta$ denotes the sum-rate requirement. 
	\end{remark}
	
	\begin{remark}
		Let ${\bf bd}\,{\cal R}_\text{C-S}$ denote the set of all the boundary points of the C-S region. Then $(S_{\min}, R_{\rm T})$ and $(S_{\rm I}, R_{\max})$ are the two critical points of ${\bf bd}\,{\cal R}_\text{C-S}$, where $R_{\rm T}$ is the optimal value of Problem ${\bf P_3}$ under $\delta = S_{\min}$ and $S_{\rm I}$ is the optimal value of Problem ${\bf P_4}$ under $\zeta = R_{\max}$. All the points of ${\bf bd}\,{\cal R}_\text{C-S}$ between $(S_{\min}, R_{\rm T})$ and $(S_{\rm I}, R_{\max})$ can be obtained by solving  Problem ${\bf P_3}$ (Problem ${\bf P_4}$) subject to an assigned parameter $\delta \geq S_{\min}$ ($\zeta \leq R_{\max}$), where the C-S performance trade-off occurs due to finite resources. The union of the set $\{(S,R_{\rm max})\mid S>S_I\}$ and the set $\{(S_{\rm min},R)\mid R<R_T\}$ contains all the other points of ${\bf bd}\,{\cal R}_{C-S}$.
	\end{remark}
	As mentioned in Remark 2, to determine the set ${\cal R}_\text{C-S}$, we need to practically solve Problem ${\bf P_1}$, Problem ${\bf P_2}$ and Problem ${\bf P_3}$ (Problem ${\bf P_4}$) after finishing the pilot allocation and target-AP pairing. The schemes for pilot allocation and target-AP pairing are presented in the next section, and all the algorithms needed for determining ${\cal R}_\text{C-S}$ will be presented in Section IV.
	
	 \vspace{-1em}
	\section{Pilot Allocation Scheme and Target-AP Pairing Scheme }
	
	Because the {\it uplink pilot training} depends on the pilot allocation scheme, i.e., $\{ \mathcal{P}_k \}$, while the {\it downlink information transmission and target sensing} depends on the target-AP pairing scheme, i.e., $\{ \mathcal{T}_m \}$, a pilot allocation scheme and a target-AP pairing scheme are respectively presented in the following two subsections.
	
	 \vspace{-1em}
	\subsection {\it Pilot Allocation Scheme}
	
	As mentioned above, the number of IUs is greater than the length of the pilot sequence, i.e., $K > \tau_{\rm p} $, which makes the pilot contamination is unavoidable. In order to mitigate the pilot contamination, the IUs far away from each other will reuse the same pilot sequence. To this end, a pilot allocation scheme based on the hierarchical agglomerative clustering (HAC) method is designed. First of all, each IU forms one cluster denoted by $\{ {\cal{C}}_i  \}$, thus forming $K$ one-member clusters. Denote the inter-cluster distance between $\mathcal{C}_i$ and $\mathcal{C}_j$ ($j \neq i $) by
	\begin{flalign*}
		d_{ij}  = \mathop{\min}\limits_{k \in \mathcal{C}_i , k' \in \mathcal{C}_j } \left\{  \sqrt{ (x_{{\rm U}_k}- x_{ {\rm U}_{k'}}    )^2 + (y_{{\rm U}_k}- y_{ {\rm U}_{k'}}    )^2  }     \right\}. 
	\end{flalign*}
	Then, two clusters with the maximum inter-cluster distance are merged into one cluster until  $\tau_{\rm p}$ clusters survive. Note that if $k \in \mathcal{C}_i$, it holds that $\mathcal{P}_k =  \mathcal{C}_i$. The proposed clustering scheme is summarized in Algorithm 1. 
	\begin{algorithm}[h]
		\caption{The proposed HAC algorithm for pilot allocation scheme } \label{alg:3}
		\textbf{Initialization}: \\
		~Treat each IU as a cluster head and set ${ num}= K$;\\
		~Calculate the inter-cluster distance set $\{  d_{ij} \}$;\\
		\While{~\rm ${ num} > {\tau_{\rm p}}$ }{
			~Merge the two clusters with maximum $d_{ij}$ into one cluster;  \\
			~Update ${num} := num-1$;		\\
			~Update the inter-cluster distnace set $\{  d_{ij} \}$;  \\
		}
		~Assign ${\tau_{\rm p}}$ a distinct orthogonal pilot sequence to each of ${\tau_{\rm p}}$ IU clusters.
	\end{algorithm}
    
	 \vspace{-2em}
	\subsection {\it Target-AP Pairing Scheme}
	As mentioned above, the probing range of each AP is bounded by the power budget. Thus, to obtain a good sensing performance, the AP responsible for sensing the assigned target should be close to the target. However, the exact coordinate of each target is unknown. Therefore,  the prior knowledge of the target-AP distances, i.e., $ \{ \bar{d}_{{\rm T}_t, {\rm A}_m}\}$ can be employed for target-AP pairing.
	One intuitive way of the target-AP pairing is to minimize the sum distance between targets and their serving APs. All the APs and targets in the considered system can be treated as a weighted complete bipartite graph $G = (\mathcal{M}, \mathcal{T})$, where $\{ -\bar{d}_{{\rm T}_t, {\rm A}_m}\}$ is the corresponding edge weight between ${T}_t$ and ${A}_m$. Then, the sum distance minimization problem is equivalent to optimal matching problem on $G$, which can be solved by the Kuhn–Munkres Algorithm \cite{km}.
	
	 \vspace{-1em}
	\section{Proposed C-S Region Determination}
	
	With the pilot allocation scheme and the target-AP pairing scheme, Problems ${\bf P_1}$, ${\bf P_2}$ and ${\bf P_3 }$ need to be respectively solved to obtain the C-S region.
		
	 \vspace{-1em}
	\subsection {\it Sensing-only Case}

	It is observed that ${\bf Z}_m$ and ${\bf Z}_{m'}$ ($m' \neq m$) are uncoupled in the convex constraints of Problem ${\bf P_1}$ and the objective function of Problem ${\bf P_1}$ is also convex. Thus, it can be decomposed into $M_{\rm T}$ convex sub-problems, which can be solved by CVX solvers, e.g., {\tt {SDPT3}},  in a parallel fashion.
	
	 \vspace{-1em}
	\subsection {\it Communication-only Case}
	
	From Problem ${\bf P_{2}}$, it can be observed that $\{ {\bf w}_{mk} \}_{k}$ and $\{ {\bf w}_{m'k} \}_{k}$ ($m' \neq m$) are uncoupled in the constraints. Therefore, solving Problem ${\bf P_2}$ is equivalent to solving $M$ sub-problems, of which the $m$-th sub-problem is given by
	\begin{subequations}
		\begin{align}
			{\bf P_{2.1}:} &\mathop {\max }\limits_{\left\{ {\bf w}_{mk}    \right\}_k}
			\sum_{k \in \mathcal{K}} r_k \left(  \left\{   {\bf w}_{mk}, {\bf 0}  \right\} \right) \\
			&{\rm s.t.}~ \sum_{k \in \mathcal{K}} {\bf w}_{mk}^H {\bf w}_{mk} \leq 1.
		\end{align}
	\end{subequations}
	Problem ${\bf P_{2.1}}$  can be optimally solved based on the idea of Lagrangian dual transform \cite{lar1} and quadratic fractional transform \cite{ft}. Through the Lagrangian dual transform, Problem ${\bf P_{2.1}}$ can be decomposed as an inner optimization problem and an outer optimization problem. With the closed-form optimal solution to the inner optimization problem and the KKT condition, the outer optimization problem can be reformulated. Through the quadratic fractional transform, the variables of the reformulated outer problem are decoupled and can be optimized in a BCD manner. The details are presented as follows. 
	
	By introducing auxiliary variables ${\bm \gamma} = [\gamma_1, \gamma_2,...,\gamma_K]^T$, Problem ${\bf P_{2.1}}$ can be re-written as
	\begin{subequations}
		\begin{align}
			{\bf P_{2.2}:} &\mathop {\max }\limits_{{\bm \gamma },   \left\{  {\bf w}_{mk}    \right\}_{k} }
			\sum_{k \in \mathcal{K}} \log_2 \left( 1 + \gamma_k  \right) \\
			&{\rm s.t.}~ \sum_{k \in \mathcal{K}} {\bf w}_{mk}^H {\bf w}_{mk} \leq 1, \\
			&~~~~~  \gamma_k \leq  \frac{\left|\sum_{m \in \mathcal{M}} \sqrt{p_m} \widehat{\bf h}_{mk}^H {\bf w}_{mk} \right|^2 }  { A_k\left( \left\{   {\bf w}_{mk}, {\bf 0}   \right\}   \right)   }      ,~\forall k \in \mathcal{K}.
		\end{align}
	\end{subequations}
	Similar to \cite{lar2}, with fixed $ \{ {\bf w}_{mk} \}_k $, the inner optimization problem is expressed as 
	\begin{subequations}
		\begin{align}
			{\bf P_{2.3}:} &\mathop {\max }\limits_{{\bm \gamma } }
			\sum_{k \in \mathcal{K}} \log_2 \left( 1 + \gamma_k  \right) \\
			&{\rm s.t.}~
			 \gamma_k \leq  \frac{\left|\sum_{m \in \mathcal{M}} \sqrt{p_m} \widehat{\bf h}_{mk}^H {\bf w}_{mk} \right|^2 }  { A_k\left( \left\{   {\bf w}_{mk}, {\bf 0}   \right\}   \right)   }      ,~\forall k \in \mathcal{K} \label{p_1.2_c},
		\end{align}
	\end{subequations}
	and the closed-form solution to Problem ${\bf P_{2.3}}$, denoted by ${\bm \gamma}^\star = [\gamma_1^\star, \gamma_2^\star,..., \gamma_K^\star]^T$, is given by
	\begin{flalign}
		\gamma_k^\star = \frac{\left|\sum_{m \in \mathcal{M}} \sqrt{p_m} \widehat{\bf h}_{mk}^H {\bf w}_{mk} \right|^2 }  { A_k\left( \left\{   {\bf w}_{mk}, {\bf 0}    \right\}   \right)   }. \label{gamma_k}
	\end{flalign}
	Since Problem ${\bf P_{2.3}}$ is convex, the Slater condition holds, and hence strong duality also holds. Let ${\bm \lambda} = [\lambda_1, \lambda_2,..., \lambda_K ]^T $ be the Lagrangian multiplier associated with constraint (\ref{p_1.2_c}), the Lagrangian function of Problem ${\bf {P}_{2.3} }$ is given by   
	\begin{flalign}
		&\mathcal{L}_1  ( {\bm \gamma}, {\bm \lambda}  ) = 
		\sum_{k \in \mathcal{K}}  \log_2 (1 + \gamma_k )  
		 + \sum_{k \in \mathcal{K} } \lambda_k \left( \frac{\left|\sum_{m \in \mathcal{M}} \sqrt{p_m} \widehat{\bf h}_{mk}^H {\bf w}_{mk} \right|^2 }  { A_k\left( \left\{   {\bf w}_{mk}, {\bf 0}   \right\}   \right)   } - \gamma_k   \right).  \label{lar1}
	\end{flalign}
	Denote ${\bm \lambda}^\star = [\lambda_1^\star, \lambda_2^\star,..., \lambda_K^\star]^T$ by the optimal solution to the dual problem of Problem ${\bf P_{2.3}}$. According to the KKT condition, it holds that
	\begin{flalign}
		&\left. \frac{ \partial \mathcal{L}_1  ( {\bm \gamma}, {\bm \lambda}  )  } { \partial \gamma_k } \right|_{ {\bm \gamma} = {\bm \gamma}^\star, {\bm \lambda} = {\bm \lambda}^\star   } = 0 \nonumber   \Rightarrow \nonumber \\
		&\lambda_k^\star  = \frac{1}{(1+\gamma_k^\star) \ln 2  } 
		 = \frac{A_k\left( \left\{   {\bf w}_{mk}, {\bf 0}    \right\}   \right)}{  \left(A_k\left( \left\{   {\bf w}_{mk}, {\bf 0}    \right\}   \right)+ \left|\sum_{m \in \mathcal{M}} \sqrt{p_m} \widehat{\bf h}_{mk}^H {\bf w}_{mk} \right|^2 \right)\ln 2  }. \label{gamma}
	\end{flalign}
	By substituting (\ref{gamma}) into (\ref{lar1}), Problem ${\bf P_{2.2}}$ can be re-expressed as 
	\begin{subequations}
		\begin{align}
			{\bf P_{2.4}:} &\mathop {\max }\limits_{{\bm \gamma },   \left\{  {\bf w}_{mk}    \right\}_k }
			f_1 \left(  {\bm \gamma },   \left\{  {\bf w}_{mk}    \right\}_k  \right) \\
			&{\rm s.t.}~ \sum_{k \in \mathcal{K}} {\bf w}_{mk}^H {\bf w}_{mk} \leq 1, 
		\end{align}
	\end{subequations}
	where $f_1 (  {\bm \gamma },   \left\{  {\bf w}_{mk}    \right\}_k )$ is given by (\ref{f_1}).
	\begin{figure*}
		\begin{flalign}
			&f_1 \left(  {\bm \gamma },   \left\{  {\bf w}_{mk}    \right\}_k  \right) =  \sum_{k \in \mathcal{K}}\left(  \log_2 (1 + \gamma_k ) - \frac{\gamma_k}{{\ln 2}}   + \frac{1}{{\ln 2}} \frac{  (1+\gamma_k)  \left|\sum_{m \in \mathcal{M}} \sqrt{p_m} \widehat{\bf h}_{mk}^H {\bf w}_{mk} \right|^2 }  {  A_k\left( \left\{   {\bf w}_{mk}, {\bf 0}    \right\}   \right)+ \left|\sum_{m \in \mathcal{M}} \sqrt{p_m} \widehat{\bf h}_{mk}^H {\bf w}_{mk} \right|^2  }   \right)
			\label{f_1}
		\end{flalign}
		\hrule
	\end{figure*}
	Problem ${\bf P_{2.4}}$ is non-convex due to the non-concave third term in 	$f_1 (  {\bm \gamma },   \left\{  {\bf w}_{mk}    \right\}_k  )$, which can be further handled via the following lemma based on the quadratic fractional transform.
	\begin{Lemma}
		Given a feasible set $ \mathcal{X} \subseteq \mathbb{C}^n $,  a function $F( {\bf x} ): \mathbb{C}^n  \rightarrow  \mathbb{C}$ and a function $ G({\bf x}):  \mathbb{C}^n \rightarrow \mathbb{R}^{+}  $. Then, the following two problems are equivalent.
			\begin{align*}
				{\bf P_{5}:} &\mathop {\max }\limits_{ {\bf x} }
				   ~ \mu_1 \left({\bf x}  \right)  \triangleq \frac{  |F\left(  {\bf x} \right)|^2  } {  G\left(    {\bf x}  \right)   }  \nonumber  \\
				&{\rm s.t.}~  {\bf x} \in \mathcal{X}. \nonumber  \\
				{\bf P_{6}:} &\mathop {\max }\limits_{ {\bf x}, y }
				  ~ \mu_2 \left({\bf x},y  \right)  \triangleq   2 {\rm Re}\{ y^* F\left( {\bf x}  \right)\} - |y|^2 G({\bf x})  \nonumber  \\
				&{\rm s.t.}~  {\bf x} \in \mathcal{X}, y \in \mathbb{C}. \nonumber 
			\end{align*}
	\end{Lemma}
\begin{IEEEproof}
	Observe that Problem ${\bf P_6}$ is concave  w.r.t. the unconstrained variable $y$. Therefore, with given ${\bf x}$, the optimal $y^\star$ satisfies that $\partial \mu_2 \left({\bf x},y  \right) / \partial y = 0$, which indicates that $y^\star = F( {\bf x} ) / G({\bf x} ) $. By substituting $y^\star$ into $\mu_2 \left({\bf x},y  \right)$, it holds that $\mu_1 \left({\bf x}  \right) = \mu_2 \left({\bf x},y^\star  \right)$. Then, Problems ${\bf P_5}$ and ${\bf P_6}$ are equivalent.
\end{IEEEproof}

	With auxiliary variables ${\bf g} \triangleq [g_1, g_2,...,g_K]^T $ and by Lemma 1, Problem ${\bf P_{2.4}}$ can be rewritten as the following problem
	\begin{subequations}
		\begin{align}
			{\bf P_{2.5}:} &\mathop {\max }\limits_{{\bm \gamma },   \left\{  {\bf w}_{mk}    \right\}_k, {\bf g} }
			f_2 \left(  {\bm \gamma },   \left\{  {\bf w}_{mk}    \right\}_k, {\bf g}  \right) \\
			&{\rm s.t.}~ \sum_{k \in \mathcal{K}} {\bf w}_{mk}^H {\bf w}_{mk} \leq 1, 
		\end{align}
	\end{subequations}
	where  $f_2 \left(  {\bm \gamma },   \left\{  {\bf w}_{mk}    \right\}_k, {\bf g}  \right)$ is given by (\ref{f_2}).
	\begin{figure*}
		\begin{flalign}
			f_2 \left(  {\bm \gamma },   \left\{  {\bf w}_{mk}    \right\}_k, {\bf g}  \right) = & \sum_{k \in \mathcal{K}}\left(\log_2 (1 + \gamma_k )-  \frac{\gamma_k }{\ln 2  }   +  \frac{2}{\ln 2  }   {\rm Re} \left\{  \sqrt{(1+\gamma_k)}  g_k^*  {    \sum_{m \in \mathcal{M}} \sqrt{p_m} \widehat{\bf h}_{mk}^H {\bf w}_{mk}  } \right\} \right)   \nonumber \\
			&~~~
			-  \sum_{k \in \mathcal{K} } \frac{\left|  g_k \right|^2}{\ln 2  }  {  \left(A_k\left( \left\{   {\bf w}_{mk}, {\bf 0}    \right\}   \right)  + \left|\sum_{m \in \mathcal{M}} \sqrt{p_m} \widehat{\bf h}_{mk}^H {\bf w}_{mk} \right|^2 \right)  }  
			\label{f_2}
		\end{flalign}
		\hrule
	\end{figure*}
	As ${\bf g}$ is unconstrained  and $f_2 \left(  {\bm \gamma },   \left\{  {\bf w}_{mk}    \right\}_k, {\bf g}  \right)$ is concave w.r.t. ${\bf g}$, the optimal ${\bf g}^\star = [ g_1^\star, g_2^\star,..., g_K^\star ]^T$ satisfies $\partial f_2 \left(  {\bm \gamma },   \left\{  {\bf w}_{mk}  \right\}_k, {\bf g}  \right) / \partial g_k  = 0$ with given ${\bm \gamma}$ and $\{ {\bf w}_{mk} \}_k$, thereby yielding
	\begin{flalign}
		g_k^\star = \frac{\sqrt{(1+\gamma_k)} {    \sum_{m \in \mathcal{M}} \sqrt{p_m} \widehat{\bf h}_{mk}^H {\bf w}_{mk}  }}
		{A_k\left( \left\{   {\bf w}_{mk}, {\bf 0}    \right\}   \right)  + \left|\sum_{m \in \mathcal{M}} \sqrt{p_m} \widehat{\bf h}_{mk}^H {\bf w}_{mk} \right|^2 } .   \label{g_k}
	\end{flalign}
	Notice that the first term, the third term and the fourth term in $A_k\left( \left\{   {\bf w}_{mk}, {\bf 0}    \right\}   \right)$ (cf. (\ref{noise})) are quadratic w.r.t. $\{ {\bf w}_{mk} \}_k$. Besides, with given  $\{ {\bf w}_{m'k} \}_{m' \neq m}$, the second term and the fifth term in $A_k\left( \left\{   {\bf w}_{mk}, {\bf 0}    \right\}   \right)$ are linear w.r.t.  $\{ {\bf w}_{mk} \}_k$. That is, $A_k\left( \left\{   {\bf w}_{mk}, {\bf 0}    \right\}   \right)$ is convex w.r.t. $\{ {\bf w}_{mk} \}_k$ with given $\{ {\bf w}_{m'k} \}_{m' \neq m}$, implying that $f_2 \left(  {\bm \gamma },   \left\{  {\bf w}_{mk}    \right\}_k, {\bf g}  \right)$ is concave w.r.t. $\{ {\bf w}_{mk} \}_k$. As a result, ${\bm \gamma },   \left\{  {\bf w}_{mk}  \right\}_k$ and ${\bf g}$ can be optimized in a BCD manner, since they are decoupled in the constraint of Problem ${\bf P_{2.5}}$. Moreover, when other variables are fixed, the closed-form expressions of the optimal  ${\bm \gamma}^\star$ and ${\bf g}^\star$ have been obtained in (\ref{gamma_k}) and (\ref{g_k}), respectively, while $\{ {\bf w}^\star_{mk} \}_k$ is optimized by solving
	\begin{subequations}
		\begin{align}
			{\bf P_{2.6}:} &\mathop {\max }\limits_{ \left\{  {\bf w}_{mk}    \right\}_k }
			f_2 \left(  {\bm \gamma },   \left\{  {\bf w}_{mk}    \right\}_k, {\bf g}  \right) \\
			&{\rm s.t.}~ \sum_{k \in \mathcal{K}} {\bf w}_{mk}^H {\bf w}_{mk} \leq 1. 
		\end{align}
	\end{subequations}
	
	Furthermore, Problem ${\bf P_2}$ can also be solved in a parallel fashion instead of solving the $M$ sub-problems sequentially. Finally, by integrating all the preceding approaches, we end up with Algorithm 2 for handling Problem ${\bf P_2}$.

	\begin{algorithm}[h]
		\caption{The proposed BCD-based algorithm for solving Problem ${\bf P_{2}}$} \label{alg:3}
		{\textbf{Initialization}: }\\
		~Find a feasible $\{ {\bf w}_{mk} \}$ to Problem ${\bf P_{2}}$;\\
		\While{~\it the stop criterion is not satified }{
			\For{  $m \in \mathcal{M}$ {\bf parallel}  }{
				~Calculate ${\bm  \gamma }$ and ${\bf g}$ with $\{ {\bf w}_{mk} \}$ by (\ref{gamma_k}) and (\ref{g_k}), respectively;               \\
				\While{~\it the stop criterion is not satified }{
					~Obtain the optimal solution  $\{ {\bf w}_{mk}^\star \}_k$ by solving Problem  ${\bf P_{2.6}}$ with ${\bm \gamma}$, ${\bf g}$ and $\{ {\bf w}_{m'k}  \}_{m' \neq m}$;\\
					~Update ${\bm \gamma}$ with $\{ {\bf w}_{mk}^\star \}_k$ and $\{ {\bf w}_{m'k}  \}_{m' \neq m}$  by (\ref{gamma_k});\\
					~Update ${\bf g}$ with  $\{ {\bf w}_{mk}^\star \}_k$,  $\{ {\bf w}_{m'k}  \}_{m' \neq m} $ and ${\bm \gamma}$ by (\ref{g_k});\\
				}
			}
			Update $\{ {\bf w}_{mk} \} := \{  {\bf w}_{mk}^\star \}$; 
		}
	\end{algorithm}

	 \vspace{-1em}
	\subsection {\it ISAC Case} 
	
	Similar to the reformations in {\it the  communication-only case},  based on the Lagrangian dual transform and the quadratic fractional transform, the objective function (\ref{p_3_a}) of Problem ${\bf P_3}$ can be rewritten as $f_3 \left( \left\{  {\bf w}_{mk}    \right\}, \{ {\bf Z}_m \}, {\bm \iota} , {\bm \kappa}   \right)$ with auxiliary variables ${\bm \iota} \triangleq [\iota_1, \iota_2,..., \iota_K]^T$ and ${\bm \kappa} \triangleq [\kappa_1, \kappa_2,..., \kappa_K]^T$, which is given in (\ref{f_3}). 
	\begin{figure*}
		\begin{flalign}
			f_3 &\left( \left\{  {\bf w}_{mk}    \right\}, \{ {\bf Z}_m \}, {\bm \iota} , {\bm \kappa}   \right) =   \sum_{k \in \mathcal{K}}  \log_2(1 + \iota_k )  +    \sum_{k \in \mathcal{K} } \frac{2}{\ln 2  }   {\rm Re} \left\{  \sqrt{(1+\iota_k)}  \kappa_k^*  {    \sum_{m \in \mathcal{M}} \sqrt{p_m} \widehat{\bf h}_{mk}^H {\bf w}_{mk}  } \right\}  \nonumber \\
			&~~~~~~~~~~~~~~~~~~~~~~
			  -  \sum_{k \in \mathcal{K} } \frac{\left|  \kappa_k \right|^2}{\ln 2  }  {  \left(A_k\left( \left\{   {\bf w}_{mk}, {\bf Z}_m    \right\}   \right)  + \left|\sum_{m \in \mathcal{M}} \sqrt{p_m} \widehat{\bf h}_{mk}^H {\bf w}_{mk} \right|^2 \right)  } 
			  -\sum_{k \in \mathcal{K}} \frac{ \iota_k} {{\rm ln} 2 } 
			\label{f_3}
		\end{flalign}
		\hrule
	\end{figure*}  
	Since $\{\eta_m, {\bf w}_{mk}, {\bf Z}_m \}$, ${\bm \iota} $ and  ${\bm \kappa}$ are uncoupled in the feasible set, they can also be optimized in a BCD manner. Besides, the closed-form expressions of the optimal ${\bm \iota}^\star \triangleq [\iota_1^\star, \iota_2^\star,..., \iota_K^\star ]^T $ with fixed $\{ {\bf w}_{mk}, {\bf Z}_m \}$ and those of the optimal ${\bm \kappa}^\star \triangleq [\kappa_1^\star, \kappa_2^\star,..., \kappa_K^\star]^T$ with fixed $\{ {\bf w}_{mk}, {\bf Z}_m \}$ and ${\bm \iota}$ are respectively  given  by
	\begin{flalign}
		& \iota_k^\star = \frac{\left|\sum_{m \in \mathcal{M}} \sqrt{p_m} \widehat{\bf h}_{mk}^H {\bf w}_{mk} \right|^2 }  { A_k\left( \left\{   {\bf w}_{mk}, {\bf Z}_m    \right\}   \right)   } \label{iota_k}  ~{\rm and}  \\
		& \kappa_k^\star = \frac{\sqrt{(1+\iota_k)} {    \sum_{m \in \mathcal{M}} \sqrt{p_m} \widehat{\bf h}_{mk}^H {\bf w}_{mk}  }}
		{A_k\left( \left\{   {\bf w}_{mk}, {\bf Z}_m    \right\}   \right)  + \left|\sum_{m \in \mathcal{M}} \sqrt{p_m} \widehat{\bf h}_{mk}^H {\bf w}_{mk} \right|^2 } .   \label{kappa_k}
	\end{flalign}
	With given ${\bm \iota}$ and ${\bm \kappa}$, $ \{ \eta_m^\star, {\bf w}_{mk}^\star, {\bf Z}_m^\star \}$ can be obtained through solving the following problem. 
	\begin{subequations}
		\begin{align}
			{\bf P_{3.1}:} &\mathop {\max }\limits_{\left\{ \eta_m, {\bf w}_{mk}, {\bf Z}_m    \right\}}   
			f_3 \left( \left\{  {\bf w}_{mk}    \right\}, \left\{ {\bf Z}_m \right\}, {\bm \iota} , {\bm \kappa}   \right) \\
			&{\rm s.t.}~
			 {\rm (\ref{power1}), (\ref{power2}), (\ref{p_3_b}), (\ref{p_3_c}) }. \nonumber  
		\end{align}
	\end{subequations}
    It is observed that $	f_3 ( \{  {\bf w}_{mk}    \}, \{ {\bf Z}_m \}, {\bm \iota} , {\bm \kappa}  )$ is non-concave w.r.t. $\{ {\bf w}_{mk}  \}$ since $	A_k( \{   {\bf w}_{mk}, {\bf Z}_m    \}   ) $ (cf. (\ref{noise})) is non-convex w.r.t. $\{  {\bf w}_{mk} \}$.
    
    To tackle the non-concave objective function of Problem ${\bf P_{3.1}}$, i.e., $	f_3 ( \{  {\bf w}_{mk}    \}, \{ {\bf Z}_m \}, {\bm \iota} , {\bm \kappa}  )$, we define convex functions ${B}_k ( \{  {\bf w}_{mk}, {\bf Z}_m \})$ and $C_k( \{  {\bf w}_{mk} \} ) $ as (\ref{B_k}) and (\ref{C_k}), respectively. Then, $A_k( \{   {\bf w}_{mk}, {\bf Z}_m   \}  )$ can be rewritten as
		\begin{figure*}
		\begin{flalign}
			&B_k  \left( \left\{   {\bf w}_{mk}, {\bf Z}_m    \right\}   \right)  \triangleq   \sum_{k' \neq k} \left| \sum_{m \in \mathcal{M}}   \sqrt{ p_m } \widehat{\bf h}_{mk}^H {\bf w}_{mk'}   \right|^2+ \sum_{k' \in \mathcal{K}  } \sum_{m \in \mathcal{M}} p_m {\bf w}_{mk'}^H {\bf \Theta}_{mk} {\bf w}_{mk'} +1 	
			\label{B_k}  
	 \\  
			& 
			+ \sum_{m \in \mathcal{M}_{\rm T} }  p_m\left( \widehat{\bf h}_{mk}^H {\bf Z}_m \widehat{\bf h}_{mk} + {\rm Tr} \left\{  {\bf \Theta}_{mk} {\bf Z}_m  \right\}   \right) + \sum_{k' \in \mathcal{K}} \sum_{m \in \mathcal{M}} \sum_{m' > m} \left| \sqrt{p_m} {\bf \Theta}_{mm'k}^H  {\bf w}_{mk'} +  \sqrt{p_{m'}} {\bf w}_{m'k'}      \right|^2    
			\nonumber 
		\end{flalign}
		\hrule
		\begin{flalign}
			C_k\left( \left\{   {\bf w}_{mk}   \right\}   \right) & \triangleq  \sum_{k' \in \mathcal{K}} \sum_{m \in \mathcal{M}} \sum_{m' > m} \left(
			p_m {\bf w}_{mk'}^H {\bf \Theta}_{mm'k} {\bf \Theta}_{mm'k}^H {\bf w}_{mk'} + p_{m'} {\bf w}_{m'k'}^H {\bf w}_{m'k'}
			\right)
			\label{C_k}  
		\end{flalign}
		\hrule	
	\end{figure*}
	\begin{flalign}
		&A_k\left( \left\{   {\bf w}_{mk}, {\bf Z}_m    \right\}   \right) = {B}_k \left( \{  {\bf w}_{mk}, {\bf Z}_m \} \right) - 	C_k  \left( \left\{ {\bf w}_{mk}    \right\} \right).
	\end{flalign}
	By introducing auxiliary variables ${\bf u} = [u_1, u_2,..., u_K]^T $ in which $u_k = C_k( \{  {\bf w}_{mk} \} ) $, and then, substituting $A_k  \{ {\bf w}_{mk}, {\bf Z}_m   \})$ in $f_3 ( \{  {\bf w}_{mk}   \}, \{ {\bf Z}_m \}, {\bm \iota} , {\bm \kappa} )$ by $( {B}_k ( \{  {\bf w}_{mk}, {\bf Z}_m \})-u_k)$, Problem ${\bf P_{3.1}}$ can be reformulated as 
	\begin{subequations}
		\begin{align}
			{\bf P_{3.2}:} &\mathop {\max }\limits_{\left\{ \eta_m, {\bf w}_{mk}, {\bf Z}_m    \right\}, {\bf u}  }   
			f_4 \left( \left\{  {\bf w}_{mk}    \right\}, \{ {\bf Z}_m \}, {\bm \iota} , {\bm \kappa}, {\bf u}   \right) \\
			&{\rm s.t.} ~ u_k \leq  	C_k\left( \left\{   {\bf w}_{mk}   \right\}   \right),  \forall k \in \mathcal{K},  \label{p_3_2_b} \\
			&~~~~~  {\rm (\ref{power1}), (\ref{power2}), (\ref{p_3_b}), (\ref{p_3_c}) }, \nonumber  
		\end{align}
	\end{subequations}
	where $f_4 \left( \left\{  {\bf w}_{mk}    \right\}, \{ {\bf Z}_m \}, {\bm \iota} , {\bm \kappa}, {\bf u}   \right)$ is a concave function w.r.t. $\{ {\bf w}_{mk}  \}$ given in (\ref{f_4}).
	\begin{figure*}
		\begin{flalign}
			f_4 &\left( \left\{  {\bf w}_{mk}    \right\}, \{ {\bf Z}_m \}, {\bm \iota} , {\bm \kappa} , {\bf u}   \right) =   \sum_{k \in \mathcal{K}}  \log_2(1 + \iota_k )  +    \sum_{k \in \mathcal{K} } \frac{2}{\ln 2  }   {\rm Re} \left\{  \sqrt{(1+\iota_k)}  \kappa_k^*  {    \sum_{m \in \mathcal{M}} \sqrt{p_m} \widehat{\bf h}_{mk}^H {\bf w}_{mk}  } \right\}  \nonumber \\
			&~~~~~~~~~~
			-  \sum_{k \in \mathcal{K} } \frac{\left|  \kappa_k \right|^2}{\ln 2  }  {  \left(B_k\left( \left\{   {\bf w}_{mk}, {\bf Z}_m    \right\}   \right) - u_k + \left|\sum_{m \in \mathcal{M}} \sqrt{p_m} \widehat{\bf h}_{mk}^H {\bf w}_{mk} \right|^2 \right)  } -\sum_{k \in \mathcal{K}} \frac{ \iota_k} {\ln 2 }  
			\label{f_4}
		\end{flalign}
		\hrule
	\end{figure*} 
	However, the constraints (\ref{power2}) and (\rm \ref{p_3_b}) are still non-convex.
	
	To tackle the non-convex constraint  (\rm \ref{p_3_b}), define $D_{mn} ( \eta_m, \{ {\bf w}_{mk} \}, {\bf Z}_m  ) $ as (\ref{D_mn}) and introduce auxiliary variables  ${\bm \xi} = [\xi_{11}, \xi_{12},..., \xi_{M_{\rm T} N}]^T$ in which $\xi_{mn} = |  D_{mn} ( \eta_m, \{ {\bf w}_{mk} \}, {\bf Z}_m  ) |$.
	\begin{figure*}[t]
		\begin{flalign}
			D_{mn} \left( \eta_m ,  \left\{ {\bf w}_{mk}   \right\}, {\bf Z}_m      \right) \triangleq  \eta_m   \widetilde{P}_m (\bar{\theta}_n)-  p_m {\bf a}^H(\bar{\theta}_n) \left( \sum_{k \in \mathcal{K} } {\bf w}_{mk} {\bf w}_{mk}^H + {\bf Z}_m   \right)  {\bf a}( \bar{\theta}_n ) \label{D_mn}
		\end{flalign}
		\hrule
	\end{figure*}
	Then, (\rm \ref{p_3_b}) can be re-expressed as
\begin{subequations}
\begin{flalign}
	  \xi_{mn} &\geq    D_{mn} \left(\eta_m , \left\{ {\bf w}_{mk} \right\}, {\bf Z}_m         \right),  \label{mse1} \\
	  -\xi_{mn} &\leq    D_{mn} \left(\eta_m , \left\{ {\bf w}_{mk} \right\}, {\bf Z}_m         \right),  \label{mse2} \\
	{\rm and }  &\sum_{m \in \mathcal{M}_{\rm T}} \sum_{n \in \mathcal{N}} \frac{\xi_{mn}^2}{M_{\rm T}  N} \leq \delta. \label{mse3}
\end{flalign}
\end{subequations}

As for (\ref{power2}), by introducing an auxiliary variable $v<1$ satisfying 
\begin{flalign}
	\sum_{k \in \mathcal{K} } {\bf w}_{mk}^H {\bf w}_{mk} + {\rm Tr}\{ {\bf Z}_m \} \geq v, \forall m \in \mathcal{M}_{\rm T},  \label{p_3_3_c}
\end{flalign}
and embedding a penalty term $(1-v)$ into the objective function, (\ref{power2})  can be relaxed as follows:
\begin{flalign}
	\sum_{k \in \mathcal{K} } {\bf w}_{mk}^H {\bf w}_{mk} + {\rm Tr}\{ {\bf Z}_m \} \leq  1, \forall m \in \mathcal{M}_{\rm T},    \label{p_3_3_b}
\end{flalign}
As a result, Problem ${\bf {P}_{3.2} }$ is equivalent to the following problem 
	\begin{subequations}
		\begin{align}
			{\bf P_{3.3}:} &\mathop {\max }\limits_{\left\{ \eta_m, {\bf w}_{mk}, {\bf Z}_m    \right\}, {\bf u}, {\bm \xi}, v  }   
			f_4 \left( \left\{  {\bf w}_{mk}    \right\}, \{ {\bf Z}_m \}, {\bm \iota} , {\bm \kappa}, {\bf u}   \right) 
			 -  q(1-v)  \\
			&{\rm s.t.}~  {\rm (\ref{power1}), (\ref{p_3_c}), ({\ref{p_3_2_b}}), (\ref{mse1}), (\ref{mse2}), (\ref{mse3}), (\ref{p_3_3_c}), (\ref{p_3_3_b}) }, \nonumber  
		\end{align}
	\end{subequations}  
where $q$ is a positive hyper-parameter in order to enforce  $\sum_{k \in \mathcal{K} } {\bf w}_{mk}^H {\bf w}_{mk} + {\rm Tr}\{ {\bf Z}_m \} = 1$ for the optimal solution to Problem ${\bf {P}_{3.3} }$.

	Problem ${\bf P_{3.3} }$ is still non-convex due to the non-convex constraints (\ref{p_3_2_b}), (\ref{mse1}) and (\ref{p_3_3_c}). Nevertheless, one can find that both sides of (\ref{p_3_2_b}), (\ref{mse1}) and (\ref{p_3_3_c}) are convex functions. Therefore, by their first-order approximations, we can come up with the following restricted convex approximation problem:
	\begin{subequations}
		\begin{align}
			{\bf P_{3.4}:} &\mathop {\max }\limits_{\left\{ \eta_m, {\bf w}_{mk}, {\bf Z}_m    \right\}, {\bf u}, {\bm \xi}, v  }   
			f_4 \left( \left\{  {\bf w}_{mk}    \right\}, \{ {\bf Z}_m \}, {\bm \iota} , {\bm \kappa}, {\bf u}   \right) 
			 -  q(1-v) \\
			&{\rm s.t.} ~   u_k \leq \widetilde{C}_k \left( \left\{   {\bf w}_{mk},  \bar{\bf w}_{mk}  \right\}   \right)  ,  \\
			& ~~~~~   \xi_{mn}  \geq \widetilde{D}_{mn} \left(  \eta_m, \left\{  {\bf w}_{mk}, \bar{\bf w}_{mk} \right\}, {\bf Z}_m     \right) ,      \\
			&~~~~~   	\sum_{k \in \mathcal{K}} \left(  \bar{ \bf w }_{mk}^H \bar{\bf w}_{mk} +   2 {\rm Re} \left\{  \bar{\bf w}_{mk}^H  \left(  {\bf w}_{mk} - \bar{\bf w}_{mk}     \right)       \right\}  \right)   
			+ {\rm Tr} \left\{  {\bf Z}_m  \right\}        \geq v  ,                   \\	
			&~~~~~  {\rm (\ref{power1}), (\ref{p_3_c}), (\ref{mse2}), (\ref{mse3}),  (\ref{p_3_3_b})  }, \forall k \in \mathcal{K},  \forall m \in \mathcal{M}_{\rm T},  n \in \mathcal{N},  \nonumber  
		\end{align}
	\end{subequations}
	where $\{ \bar{\bf w}_{mk}\}$ is feasible to Problem ${\bf P_{3.3}}$, and $\widetilde{C}_k \left( \left\{   {\bf w}_{mk},  \bar{\bf w}_{mk}  \right\}   \right)$ and $\widetilde{D}_{mn} (  \eta_m,\{  {\bf w}_{mk}, \bar{\bf w}_{mk}\}, {\bf Z}_m    ) $ are given in (\ref{C_k_bar}) and (\ref{D_mn_bar}), respectively.
	\begin{figure*}
		\begin{flalign}
			&\widetilde{C}_k \left( \left\{   {\bf w}_{mk},  \bar{\bf w}_{mk}  \right\}   \right) \triangleq    \sum_{k' \in \mathcal{K}} \sum_{m \in \mathcal{M}} \sum_{m' > m} \left(
			p_m \bar{\bf w}_{mk'}^H {\bf \Theta}_{mm'k} {\bf \Theta}_{mm'k}^H  \bar{\bf w}_{mk'} + p_{m'} \bar{\bf w}_{m'k'}^H \bar{\bf w}_{m'k'} \right. \nonumber \\
			&~~~~ \left.
			+ 2 {\rm Re} \left\{  p_m \bar{\bf w}_{mk'}^H {\bf \Theta}_{mm'k} {\bf \Theta}_{mm'k}^H \left( {\bf w}_{mk'}- \bar{\bf w}_{mk'}   \right) + p_{m'} \bar{\bf w}_{m'k'}^H \left(  {\bf w}_{m'k'} - \bar{\bf w}_{m'k'}   \right) \right\}
			\right)
			\label{C_k_bar} 
		\end{flalign}
		\hrule
		
		\begin{flalign}
			&\widetilde{D}_{mn} \left( \eta_m ,  \left\{ {\bf w}_{mk}, \bar{\bf w}_{mk}   \right\}, {\bf Z}_m \right) \triangleq \nonumber \\
			&~~~~ \eta_m   \widetilde{P}_m (\bar{\theta}_n)-  p_m {\bf a}^H(\bar{\theta}_n) \left( \sum_{k \in \mathcal{K} } \left(\bar{\bf w}_{mk} \bar{\bf w}_{mk}^H + 2 {\rm Re} \left\{ \bar{\bf w}_{mk} ( {\bf w}_{mk} - \bar{\bf w}_{mk}  )^H     \right\}     \right)+ {\bf Z}_m   \right)  {\bf a}( \bar{\theta}_n )
			\label{D_mn_bar}  
		\end{flalign}
		\hrule
	\end{figure*}
    Then one can apply the SCA method \cite{sca} to Problem ${\bf P_{3.4}}$ in order to get a good sub-optimal solution to Problem ${\bf P_3}$, thereby yielding the proposed BCD-SCA based algorithm (termed Algorithm 3) for this end.
	\begin{algorithm}[!h]
		\caption{The proposed BCD-SCA based algorithm for solving Problem ${\bf P_{3}}$} \label{alg:3}
		{\textbf{Initialization}: }\\
		~Find a feasible $\{ \bar{\bf w}_{mk}, \bar{\bf Z}_m \}$ to Problem ${\bf P_{3.3}}$;\\
		~Calculate ${\bm \iota}$ and ${\bm \kappa}$  with $\{ \bar{\bf w}_{mk}, \bar{\bf Z}_m  \}$ by (\ref{iota_k}) and (\ref{kappa_k}), respectively;    \\
		\While{~\it the stop criterion is not satified }{
			\While{~\it the stop criterion is not satified }{
				~Obtain the optimal solution $\{ {\bf w}_{mk}^\star, {\bf Z}_m^\star  \}$  by solving Problem ${\bf P_{3.4}}$ with ${\bm \iota}$, ${\bm \kappa}$ and $\{ \bar{\bf w}_{mk} \}$; \\
				~Update $\{ \bar{\bf w}_{mk}   \} :=  \{  {\bf w}_{mk}^\star  \};$
				
			}
			~Update ${\bm \iota}$ with $\{ {\bf w}_{mk}^\star, {\bf Z}_m^\star \}$ by (\ref{iota_k}); \\
			~Update ${\bm \kappa}$ with $\{{\bf w}_{mk}^\star, {\bf Z}_m^\star   \}$ and ${\bm \iota} $ by (\ref{kappa_k}); 			
		}
	\end{algorithm}

	 \vspace{-1em}
	\section{Numerical Results}
	In this section, some numerical results are presented to show the proposed C-S region of the Cell-Free massive MIMO ISAC system, i.e., the set ${\cal R}_\text{C-S}$ defined in (\ref{c-s}), as well as its characteristics, insights and impacts of system parameters on the C-S performance trade-off. The system parameters used in the obtained numerical results are listed in Table \Rmnum{1}, provided that all the APs, IUs, and targets are uniformly distributed within a square of $D\times D$ (${\rm km}^2$).


	\begin{table}
			\caption{INSTANCE SYSTEM PARAMETERS}
		\centering
		\begin{tabular}{cc|cc|cc}
			\hline
			{\bf AP Parameters}&	{\bf Values} & 	{\bf IU Parameters}&	{\bf Values}&{\bf Target Parameters}&	{\bf Values} \\
			$M$ &  $6$  &  $ K$ & $9$ &
			$T$ & $3$   \\
		    $N_{\rm T}$  & $12$   & $p_{\rm p}$ &  $150$ dB &  $\alpha_t$ & $0.8$
			 \\
			 $p_m$ & $125$  dB &  $\tau_{\rm p}$  &  $5$
		     & $\Delta \theta$  &  $10^{\circ}$
			    \\
			$D$  & $2~{\rm km}$ &  $\tau_{\rm c}$  &  $40$  & $N$ & $181$ \\
			$\sigma^2_{ {\rm T}_t, {\rm A}_m}$  & $-160$ dB  & $\sigma_{{\rm U}_k, {\rm T}_t  }^2$ & $-160$  dB &  $\chi_{{\rm T}_t, {\rm A}_m}^2 $  &  $-30$  dB  \\ 
			\hline
		\end{tabular}
	\end{table}

\begin{figure*}[!t]
	\centering
	\subfloat[The impact of the APs' number on the C-S region of the ISAC.] {\includegraphics[width=.45\textwidth]{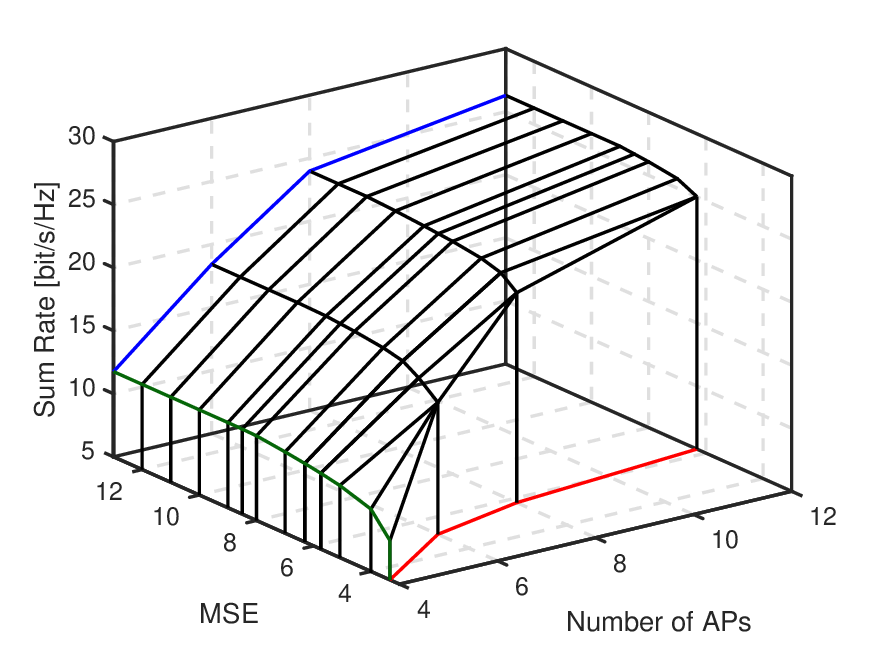}
	}
	\hfil
	\subfloat[C-S region of the ISAC when $M = 4$.]{\includegraphics[width=.45\textwidth]{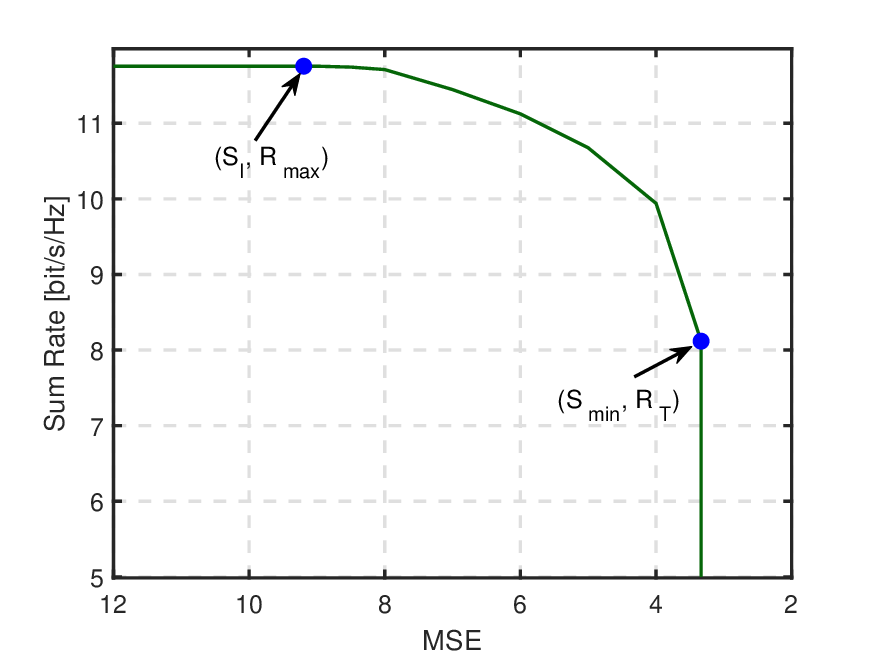}}
	\caption{The C-S region in Cell-Free massive MIMO ISAC systems.}
	\label{fig: c-s_region}
	\vspace{-2em}
\end{figure*}
     
    Figure \ref{fig: c-s_region}(a) shows a 3-dimensional (3-D) C-S region of the considered Cell-Free massive MIMO ISAC system obtained using all the algorithms proposed in Sections \Rmnum{3} and \Rmnum{4}, where the x-axis, y-axis and z-axis respectively represent the number of APs, sensing performance (in MSE)  and communication performance (in sum rate). For fairness, the total number of antennas is fixed as $48$ and evenly distributed among APs\footnote{A fixed power budget is allocated to each ISAC AP to guarantee its probing range, while the communication-only APs share the remaining available power evenly.}, i.e., $N_{\rm T} = 48/M$. One can observe that the sum rate increases with the number of APs ($M$), because the average distance between each IU and all the APs decreases, namely resulting in smaller path loss between them. In addition, the MSE also increases with $M$ due to smaller $N_{\rm T}$ used for each target sensing (with larger mainlobe width or less focused on the target).
    Nevertheless, it is by no means that the Cell-Free ISAC is inferior to the traditional ISAC, and more insights are to be discussed in Figure \ref{fig:sense} below. 
    Figure 3(b) shows the projection of the 3-D C-S region on the $(S, R)$ plane for $M=4$, where the two critical points $(S_{\rm I}, R_{\max})$ and $(S_{\min}, R_{\rm T})$
    on the boundary of the resulting 2-D C-S region of the ISAC reveal the achievable maximum sum rate $R_{\max}$  and minimum MSE $S_{\min}$, respectively. As stated in Remark 2, the C-S performance trade-off exists only on the boundary between the two critical points, depending on the given finite resource. Note that if $\delta<S_{\min}$ ($\zeta>R_{\max}$) Problem ${\bf P_3}$ (${\bf P_4}$) is infeasible.  
    
    \begin{figure*}[!t]
    	\centering
    	\subfloat[ Traditional cellular system with DOAs of targets being $(-45^{\circ}, 0^{\circ}, 45^{\circ} )$.]{\includegraphics[width=.30\textwidth]{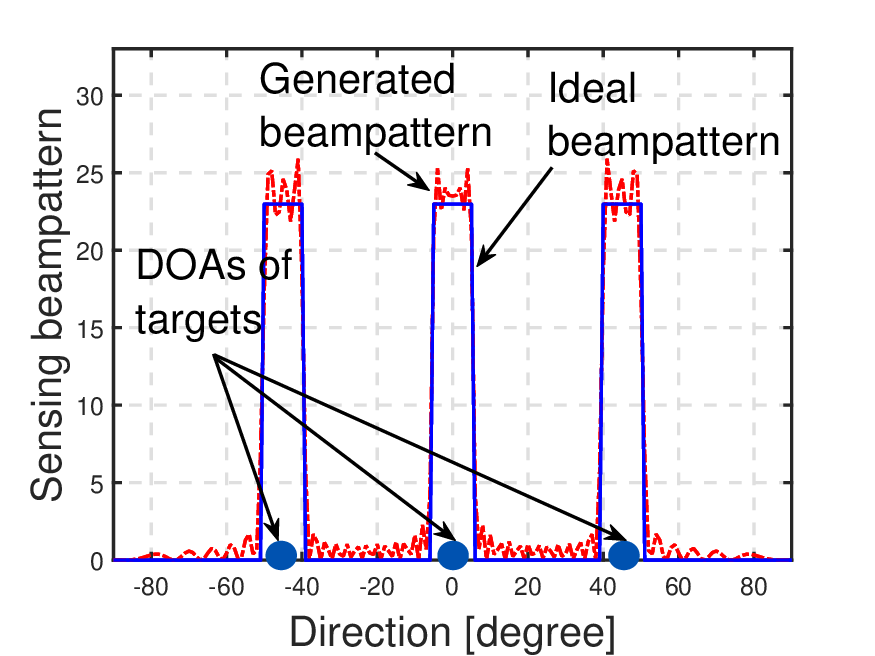}}
    	\quad
    	\hfil
    	\subfloat[Traditional cellular system with DOAs of targets being $(-80^{\circ}, 0^{\circ}, 50^{\circ})$.]{\includegraphics[width=.30\textwidth]{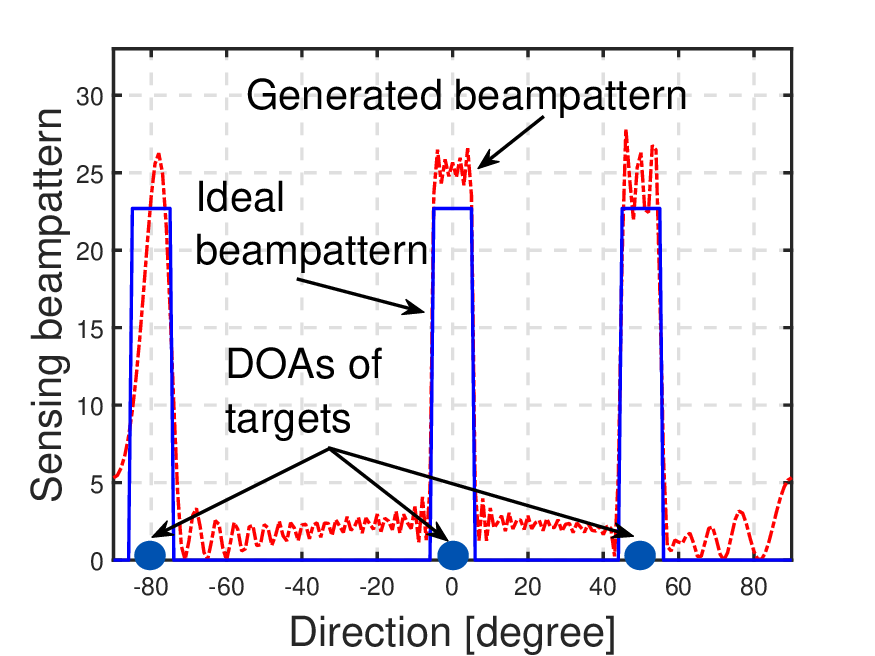}}
    	\quad
    	\subfloat[Cell-Free system (beampatterns are almost the same for each AP). ]{\includegraphics[width=.30\textwidth]{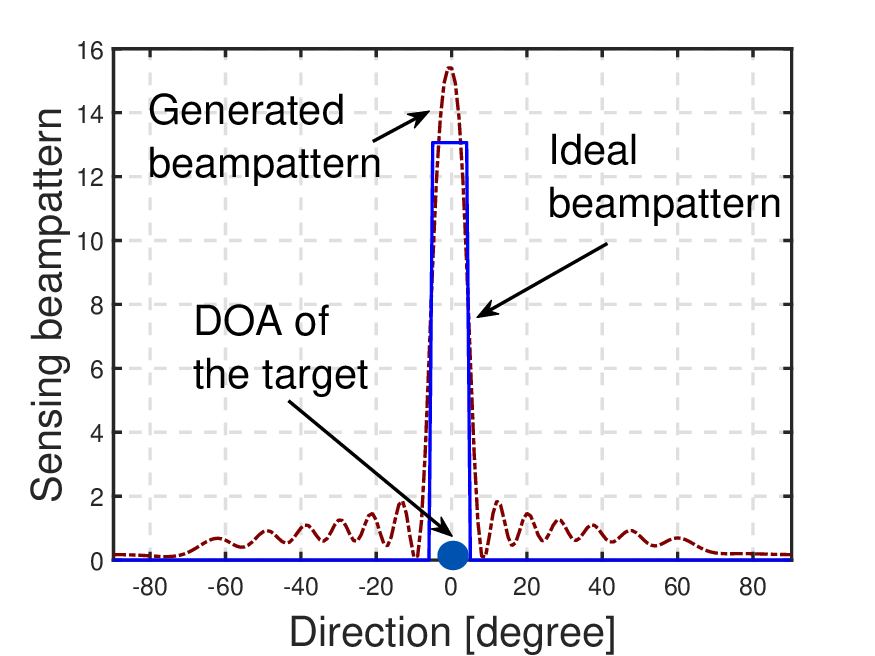}}
    	\caption{Sensing-only comparison (under $3$ targets)
    		between the traditional cellular system (one BS equipped with $48$ antennas) 
    		and the Cell-Free system ($3$ APs, each equipped with $16$ antennas).}
    	\label{fig:sense}
    	\vspace{-2em}
    \end{figure*}
 
  Let us consider the sensing-only case of $T=3$ targets for a performance comparison of the traditional system (i.e., one BS equipped with $48$ antennas)  and the Cell-Free system (with $N_{\rm T} = 16$ antennas for each of the $3$ APs). Figure \ref{fig:sense} shows the designed beampatterns for the sensing-only case, for $3$ targets with estimated DOA being $(-45^\circ,0^\circ,45^\circ)$ in Figure \ref{fig:sense}(a) and  $(-80^\circ,0^\circ,50^\circ)$ in Figure \ref{fig:sense}(b) for the traditional cellular system, while only one target with estimated DOA being $0^\circ$  in Figure \ref{fig:sense}(c) for Cell-Free system. For a fair comparison, to take the probing range into account, we consider the normalized MSE instead, which is defined as $\sum_{m \in \mathcal{M}_{\rm T}}   \mathcal{E}_m \left(  \eta_m, \{ {\bf w}_{mk} \}, {\bf Z}_m \right) / ( M_{\rm T} \eta_m^2)$. Over $2000$ realizations of the targets' positions, the obtained average normalized MSE of the centralized traditional cellular system is $0.0159$, while that of the Cell-Free system is $0.0080$ (around half of the former), thereby demonstrating much better performance for the distributed Cell-Free system. Furthermore, it can be seen from  Figure \ref{fig:sense}(a) and  Figure \ref{fig:sense}(b) that the beampattern matching quality is quite sensitive to the dispersiveness of the target distribution for the traditional cellular system, so is its sensing performance. On the contrary, in the Cell-Free system, each AP is dedicated to only one target, thus readily maintaining the target DOA being $0^\circ$ as shown in Figure  \ref{fig:sense}(c), in other words, the sensing performance of the Cell-Free system is robust against the target distribution.

\begin{figure}[t]
	\setlength{\abovecaptionskip}{0.cm}
	\begin{center}
		\centerline{\includegraphics[width=.5\textwidth]{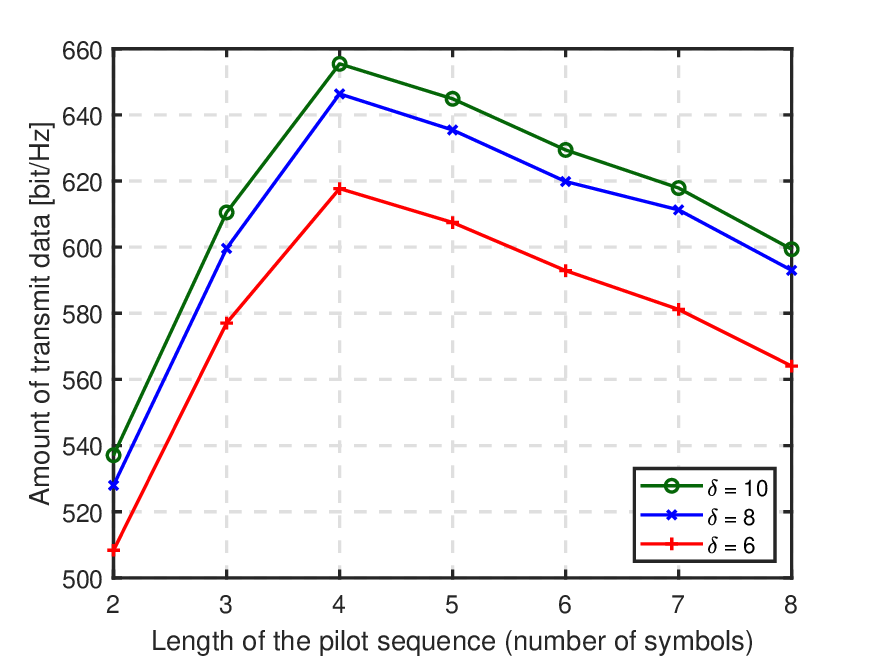}}
		\caption{The amount of transmit data versus the length of the pilot sequence.}
		\label{fig:tao_p}
	\end{center}
	\vspace{-2.5em}
\end{figure}
	Figure \ref{fig:tao_p} shows the impact of the length of the pilot sequence, i.e., $\tau_{\rm p}$, on the amount of transmit data, given by $(\tau_{\rm c}-\tau_{\rm p})	\sum_{k \in \mathcal{K}} r_k (\{   {\bf w}_{mk}, {\bf Z}_m \} ) $. It can be observed that the amount of transmit data first increases, reaches the maximum for $\tau_{\rm p} = 4$, and then  decreases with ${\tau_{\rm p}}$. The reason is because a larger $\tau_{\rm p}$ can bring a smaller channel estimation error (thus improving the downlink transmission rate) but a shorter downlink transmission time, i.e., $(\tau_{\rm c}-\tau_{\rm p})$, thus making the amount of transmit data decrease for $\tau_{\rm p}>4$ instead. Moreover, the larger the $\delta$ (i.e., lower sensing beampattern matching MSE requirement), the larger the power allocated to downlink transmission, thereby yielding more data transmitted. 
	
	\begin{figure}[t]
	\setlength{\abovecaptionskip}{0.cm}
		\begin{center}
			\centerline{\includegraphics[width=.5\textwidth]{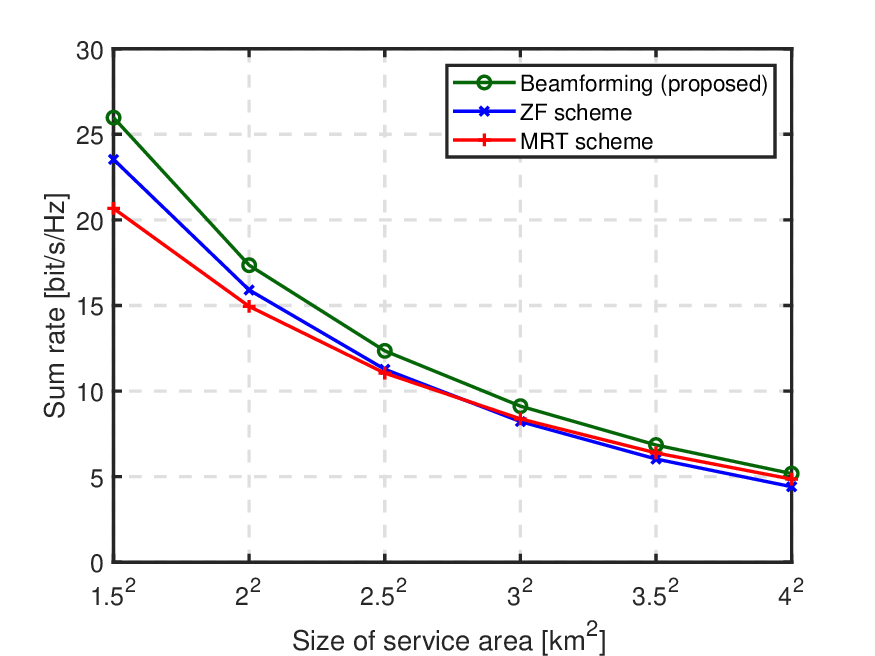}}
			\caption{Comparison among the proposed beamforming scheme, the ZF scheme and  the MRT scheme with $\delta = 6$.}
			\label{fig:zf_mrt}
		\end{center}
		\vspace{-4em}
	\end{figure}
	Figure \ref{fig:zf_mrt} shows the sum rate performance of the proposed beamforming of ISAC with $\delta = 6$ (implemented by Algorithm 3), and a comparison with the ZF scheme and the MRT scheme for different sizes of service area of the considered system. One can see from this figure that all the $3$ schemes' performances are worse for larger service area, simply because of larger average path loss incurred. Nonetheless, the proposed scheme outperforms the other two schemes, and the ZF scheme performs better (worse) than the MRT scheme for relatively small (large) service area, which is consistent with the results reported in \cite{mrt_zf_com}, as larger area corresponds to smaller SNR under the constraint on fixed power budget and noise power.

	 \vspace{-0.7em}
	\section{Conclusion}
	We have presented a novel system model and a downlink transmit beamforming design for the Cell-Free massive MIMO ISAC system, based on the first-order and second-order statistics of the MIMO channel estimation error (via the uplink pilot training) that were theoretically obtained for the first time to the best of our knowledge. The achievable  C-S region of the ISAC system under constraints of finite resources, has also been presented that can be obtained via  beamforming designs for $3$ cases (sensing-only case, communication-only case and ISAC case) using the proposed algorithms in Section \Rmnum{3} and \Rmnum{4}. Then some numerical results were provided to demonstrate the obtained C-S region and its various characteristics, specifically, the performance trade-off between communication and radar sensing on the boundary of the C-S region, the impacts of the system parameters on the beamforming performance, and advantages of the proposed beamformer beyond those used in state-of-the-art ISAC systems. Some further studies, including i) C-S performance trade-off sensitivity to various system parameters, ii) each AP sensing multiple targets, and iii) incorporation of target detection performance,  are left in the future.

	 \vspace{-1.2em}

	\appendix
	
	 \vspace{-0.9em}
	\subsection{Proof of Proposition 1}
	
	For the aggregated channel, the direct link and the cascade links are statistically independent, so are the cascade links reflected from different targets. Then, the auto-correlation matrix of ${\bf h}_{mk}$, the cross-correlation matrix of  ${\bf h}_{mk}$ and ${\bf h}_{m'k}$ ($m' \neq m$), and the cross-correlation matrix of ${\bf h}_{mk}$ and ${\bf h}_{mk'}$ ($k' \neq k$) are respectively given by 
	\begin{flalign*}
		&\mathbb{E} \left\{ {\bf h}_{mk} {\bf h}_{mk}^H   \right\}  = \mathbb{E} \left\{ {\bf h}_{{\rm U}_k, {\rm A}_m }  {\bf h}_{{\rm U}_k, {\rm A}_m }^H    \right\} 
		+      \sum_{t \in \mathcal{T}} \alpha_t^2 \mathbb{E} \left\{  | { h}_{{\rm U}_k, {\rm T}_t   } |^2  \right\} \mathbb{E}\left\{ {\bf h}_{ {\rm T}_t, {\rm A}_m  } {\bf h}_{ {\rm T}_t, {\rm A}_m}^H   \right\},  \\
		&	\mathbb{E} \left\{ {\bf h}_{mk} {\bf h}_{m'k}^H \right\} =   \mathbb{E} \left\{ {\bf h}_{{\rm U}_k, {\rm A}_m }  {\bf h}_{{\rm U}_{k}, {\rm A}_{m'} }^H \right\} 
		+\sum_{t \in \mathcal{T}} \alpha_t^2 \mathbb{E} \left\{  | { h}_{{\rm U}_k, {\rm T}_t   } |^2  \right\} \mathbb{E}\left\{ {\bf h}_{ {\rm T}_t, {\rm A}_m  } {\bf h}_{ {\rm T}_t, {\rm A}_{m'}}^H   \right\},
		 ~ {\rm and}  \\
		&
		\mathbb{E} \left\{ {\bf h}_{mk} {\bf h}_{mk'}^H   \right\}  = \mathbb{E} \left\{ {\bf h}_{{\rm U}_k, {\rm A}_m }  {\bf h}_{{\rm U}_{k'}, {\rm A}_m }^H    \right\} 
		+  \sum_{t \in \mathcal{T}} \alpha_t^2 \mathbb{E} \left\{   { h}_{{\rm U}_k, {\rm T}_t   } { h}_{{\rm U}_{k'}, {\rm T}_t   }^* \right\} \mathbb{E}\left\{ {\bf h}_{ {\rm T}_t, {\rm A}_m  } {\bf h}_{ {\rm T}_t, {\rm A}_m}^H   \right\}  .  
	\end{flalign*}
    
    According to the corresponding channel models in Section \Rmnum{2}-A, it holds that $\mathbb{E} \{ {\bf h}_{{\rm U}_k, {\rm A}_m }  \} = {\bf 0}$, $\mathbb{E} \{  {\bf h}_{{\rm U}_k, {\rm A}_m }  {\bf h}_{{\rm U}_k, {\rm A}_m }^H  \} = \beta^2(  d_{{\rm U}_k, {\rm A}_m   }) {\bf I}_{N_T}$, $\mathbb{E}\{{ h}_{{\rm U}_k, {\rm T}_t   } \} = 0$, $\mathbb{E}\{  | { h}_{{\rm U}_k, {\rm T}_t   } |^2 \} = {\beta}^2(\bar{d}_{{\rm U}_k,{\rm T}_t })+\sigma_{{\rm U}_k, {\rm T}_t  }^2$, $\mathbb{E}\{ {\bf h}_{ {\rm T}_t, {\rm A}_m  }\} = \beta(\bar{d}_{{\rm T}_t,{\rm A}_m}) \bar{ {\bf q}}_{{\rm T}_t, {\rm A}_m } $ and $\mathbb{E}\{ {\bf h}_{ {\rm T}_t, {\rm A}_m  } {\bf h}_{ {\rm T}_t, {\rm A}_m}^H\} = ( {\beta^2({  \bar{d}_{{\rm T}_t, {\rm A}_m  }})} + \sigma_{ {\rm T}_t, {\rm A}_m}^2)(  \bar{\bf q}_{{\rm T}_t, {\rm A}_m }  \bar{\bf q}_{{\rm T}_t, {\rm A}_m }^H + \chi_{{\rm T}_t, {\rm A}_m}^2 {\bf I}_{ N_{\rm T} }) $. Then, one can find that $\mathbb{E} \{ {\bf h}_{mk} {\bf h}_{mk}^H \}  =  {\bf \Phi}_{mk}$ and $\mathbb{E} \{ {\bf h}_{mk} {\bf h}_{m'k}^H \} = {\bf \Phi}_{mm'k}$. Besides, due to both $\{  {\bf h}_{{\rm U}_k, {\rm A}_m } \}$ and $\{  { h}_{{\rm U}_k, {\rm T}_t } \}$ are zero-mean and statistically independent, $\mathbb{E} \{ {\bf h}_{mk} {\bf h}_{mk'}^H \} = {\bf 0}$. Similarly, $\mathbb{E} \{ {\bf h}_{mk} {\bf h}_{m'k'}^H \}  = {\bf 0}~(m' \neq m, k' \neq k)$ can be proved.
	
	 \vspace{-1em}
	\subsection{Proof of Proposition 2}
	
	According to the definition of ${\bf C}_{mk}$ in (\ref{hat_h}),  it holds that
	\begin{flalign*}
		&\mathbb{E}\left\{ \widehat{\bf h}_{mk}   \right\} =  {\bf C}_{mk}  \mathbb{E} \left\{ {\bf y}_{{\rm p},mk}   \right\}  = {\bf 0}, ~
		{\rm and} \\
		&\mathbb{E}\left\{   \widehat{\bf h}_{mk}  \widehat{\bf h}_{mk}^H \right\}  = {\bf C}_{mk} \mathbb{E}\left\{ {\bf y}_{{\rm p},mk  }  {\bf y}_{{\rm p},mk  }^H  \right\} {\bf C}_{mk}^H  
		= 	  {\bf C}_{mk} \mathbb{E}\left\{  {\bf y}_{{\rm p},mk} {\bf h}_{mk}^H \right\}  \\
		&~~~~~~~~~~~~~~~= \sqrt{p_{\rm p} \tau_{\rm p}} {\bf C}_{mk} \mathbb{E}\left\{  {\bf h}_{mk} {\bf h}_{mk}^H \right\} 
		 =  \sqrt{p_{\rm p} \tau_{\rm p}} {\bf C}_{mk} {\bm \Phi}_{mk}.  
	\end{flalign*}
    That is, the derivation of $\mathbb{E}\{   \widehat{\bf h}_{mk}  \widehat{\bf h}_{mk}^H \}$ is equivalent to the computation of $ {\bf C}_{mk}$ which includes two parts, i.e., $\mathbb{E}\{ {\bf h}_{mk} {\bf y}_{{\rm p},mk}^H\}$ and $\mathbb{E}\{ {\bf y}_{{\rm p},mk} {\bf y}_{{\rm p},mk}^H   \}$. The former can be obtained by
    \begin{flalign*}
        \mathbb{E}\left\{ {\bf h}_{mk} {\bf y}_{{\rm p},mk}^H \right\} &= \sqrt{p_{\rm p}  \tau_{\rm p}  } \mathbb{E}\left\{ {\bf h}_{mk}  {\bf h}^H_{mk} \right\} +  \sqrt{p_{\rm p}  \tau_{\rm p}  } \sum_{k' \in \mathcal{P}_k \setminus \{ k  \}  } \mathbb{E}\left\{ {\bf h}_{mk} {\bf h}^H_{mk'} \right\} +  \mathbb{E}\left\{ {\bf h}_{mk} {\bf n}^H_{{\rm p}, mk} \right\}  \\
        & = \sqrt{p_{\rm p} \tau_{\rm p}} {\bm \Phi}_{mk},
    \end{flalign*}
    and the latter as
    \begin{flalign*}
    	\mathbb{E}\{ {\bf y}_{{\rm p},mk} {\bf y}_{{\rm p},mk}^H   \} & = p_{\rm p}  \tau_{\rm p} \sum_{k' \in \mathcal{P}_k  }\mathbb{E}\left\{  {\bf h}_{mk'} {\bf h}_{mk'}^H \right\}+ {\bf I}_{N_{\rm T} } 
    	= p_{\rm p}  \tau_{\rm p} \sum_{k' \in \mathcal{P}_k  }{\bm \Phi}_{mk'} + {\bf I}_{N_{\rm T} }.
    \end{flalign*}
    Therefore, the closed-form expression of ${\bf C}_{mk}$ is given by
	\begin{flalign*}
		{\bf C}_{mk} = \sqrt{p_{\rm p}  \tau_{\rm p}} {\bm \Phi}_{mk} \left( p_{\rm p}  \tau_{\rm p} \sum_{k' \in \mathcal{P}_k  }{\bm \Phi}_{mk'} + {\bf I}_{N_{\rm T} }  \right)^{-1}. 
	\end{flalign*}
	
	 \vspace{-1.125em}
	\subsection{Proof of Proposition 3}
	For the channel estimation error ${\bf e}_{mk}$ defined by (\ref{e}), it holds that
	\begin{flalign*}
		\mathbb{E} \left\{   {\bf e}_{mk}  \right\} = \mathbb{E} \left\{   {\bf h}_{mk}  \right\} - \mathbb{E} \left\{   \widehat{\bf h}_{mk}  \right\} = {\bf 0}.
	\end{flalign*}
	According to properties of linear MMSE method \cite{lmmse}, the channel estimation error ${\bf e}_{mk}$ and the channel estimation $\widehat{\bf h}_{mk}$ are orthogonal. Thus, the auto-correlation matrix of ${\bf e}_{mk}$ is given by
	\begin{flalign*}
		 \mathbb{E} \left\{ {\bf e}_{mk} {\bf e}_{mk}^H    \right\} &=  \mathbb{E} \left\{ \left(  {\bf h}_{mk} - \widehat{\bf h}_{mk} \right) \left(  {\bf h}_{mk} - \widehat{\bf h}_{mk} \right)^H  \right\} 
		 =\mathbb{E} \{ {\bf h}_{mk} {\bf h}_{mk}^H  \} - \mathbb{E} \{ \widehat{\bf h}_{mk} \widehat{\bf h}_{mk}^H  \}  \nonumber \\
		&= {\bm \Phi}_{mk} -  \sqrt{p_{\rm p} \tau_{\rm p}} {\bf C}_{mk} {\bm \Phi}_{mk} = {\bm \Theta}_{mk}, 
	\end{flalign*}
	and the cross-correlation matrix of ${\bf e}_{mk}$  and ${\bf e}_{m'k}$ ($m' \neq m$) is given by
	\begin{flalign*}
		&\mathbb{E}\left\{ {\bf e}_{mk} {\bf e}_{m'k}^H  \right\}   \\
		&=
		\mathbb{E}\left\{  {\bf h}_{mk}  {\bf h}_{m'k}^H   \right\} + \mathbb{E}\left\{  {\bf C}_{mk} {\bf y}_{{\rm p},mk} {\bf y}_{{\rm p},m'k}^H {\bf C}_{m'k}^H   \right\} 
		 - \mathbb{E}\left\{  {\bf h}_{mk} {\bf y}_{{\rm p},m'k}^H {\bf C}_{m'k}^H  \right\}  - \mathbb{E} \left\{  {\bf C}_{mk} {\bf y}_{{\rm p},mk} {\bf h}_{m'k}^H  \right\}  \\
		&= {\bf \Phi}_{mm'k} + p_{\rm p} \tau_{\rm p} {\bf C}_{mk} \sum_{k' \in \mathcal{P}_k  } {\bf \Phi}_{mm'k'}
		{\bf C}_{m'k}^H 
		  - \sqrt{p_{\rm p} \tau_{\rm p}  } {\bf \Phi}_{mm'k} {\bf C}_{m'k}^H- \sqrt{p_{\rm p}\tau_{\rm p}} {\bf C}_{mk} {\bf \Phi}_{mm'k} 	
		 = {\bf \Theta}_{mm'k}. 
	\end{flalign*}
	

\end{document}